\documentclass[nofootinbib,prd,aps,superscriptaddress,notitlepage]{revtex4-1}

\usepackage{hyperref}
\usepackage[utf8]{inputenc}
\usepackage{latexsym,graphicx,amssymb,amsmath} 
\usepackage{setspace,bm} 

\newcommand{\bea}{\begin{eqnarray}}
\newcommand{\eea}{\end{eqnarray}}
\newcommand{\be}{\begin{equation}}
\newcommand{\ee}{\end{equation}}

\newcommand{\rt}[1]{{}}
\newlength{\szovszel}
\newlength{\slashszel} 
\newcommand*{\sls}[1]{\mbox{%
   \settowidth{\szovszel}{\ensuremath{#1}}%
    \settowidth{\slashszel}{\ensuremath{\slash}}%
    \hspace*{0.5\szovszel}%
    \hspace*{-0.5\slashszel}%
    \slash%
    \hspace*{-0.5\szovszel}%
    \hspace*{-0.5\slashszel}%
    \ensuremath{#1}%
  }}

\begin{document}

\title{Pseudo-Goldstone excitations in chiral Yukawa-theories\\ with quadratic explicit symmetry breaking}
\author{A. Jakov\'ac}
\author{I. Kaposv\'ari}
\author{A. Patk{\'o}s}
\affiliation{Institute of Physics, E{\"o}tv{\"o}s University, H-1117 Budapest, Hungary}
\date{\today}

%\vfill
\begin{abstract}
The symmetry breakdown pattern is studied in models containing one fermion flavor multiplet and a multicomponent scalar field, supplemented with a chiral Yukawa-interaction, and in presence of an explicit symmetry breaking  source quadratic in the scalar field. In a detailed investigation of the model with $U_L(1)\times U_R(1)$ chiral symmetry it is shown that by diminishing the strength of quadratic explicit symmetry breaking one can still keep stable the mass ratio of the fermionic and the pseudo-Goldstone excitation. At the same time the mass ratio of the two bosonic excitations appears to approach a unique limiting value depending only on the infrared value of the first ratio but not on the microscopic (ultraviolet) coupling values. These observations receive a general interpretation by the existence of a slowly drifting partial fixed line located in the symmetric phase. Understanding the general conditions for its existence allows the construction of a similar theory with $U_L(2)\times U_R(2)$ chiral symmetry. All results of the present investigation were obtained with a nonperturbative functional renormalization group technique making use of the first two approximations to the gradient expansion of the effective action.      
\end{abstract}

\maketitle

\section{Introduction}

Spontaneous symmetry breaking is a central feature of elementary particle physics. The breakdown of the $SU_L(2)\times SU_R(2)$ chiral symmetry of strong interactions results in the existence of light pseudo-Goldstone pions \cite{nambu61}. Their nonzero mass is usually attributed to the presence of an explicit symmetry breaking term in the Lagrangian of the effective meson theory, linear in the meson fields. In case of electroweak interactions the would-be-Goldstone degrees of freedom transmutate into the longitudinal polarization modes of the massive vector fields without any need for explicit symmetry breaking. In this case, however, the lightness of the Higgs particle awaits some dynamical explanation, in particular, since its tree level value would be seriously altered by the quadratically divergent ultraviolet field fluctuations. By the example of the strong interactions, a popular idea is to assume that the Higgs boson itself is a pseudo-Goldstone excitation of some symmetry hiding the secret of the electroweak symmetry breaking \cite{kaplan84}. 

In the present paper, we shall investigate the effect of quantum fluctuations on the particle spectra when the explicit symmetry breaking is quadratic in the meson fields, instead of the conventional linear external field. The strength of this quadratic breaking itself is scale dependent and we will be particularly interested in investigating its effect on the renormalization evolution when its starting (microscopic) value is chosen many orders of magnitude smaller than the value of the parameters preserving the symmetry of the respective models. 

A model containing a real scalar (the "Higgs" boson) and a single fermion flavor (the "top") represents a minimal framework, which allows us to study the influence of quantum fluctuations on the Higgs potential \cite{holland04,holland05,fodor07,gerhold09,gerhold10,gies14,chu15,jakovac16,borchardt16,gies17a,gies17b}.
Though it possesses only a discrete remnant of the chiral symmetry, still an acceptable order of magnitude estimate could have been obtained for the maximal momentum value up to which the system is stable and its characteristic data (the vacuum expectation of the scalar field, the masses of the top, and of the Higgs boson) can be fixed to their experimental values.  In the present study we make use of the simplest extension of this toy model allowing us to define the Higgs as the pseudo-Goldstone boson in the broken phase of the global continuous symmetry characterizing the extension. 

The simplest extension is to replace the real scalar by a complex one and couple it to the single fermion flavor with a chiral Yukawa interaction. This model has a $U_L(1)\times U_R(1)$ global chiral symmetry, and the symmetry breaking pattern $U_L(1)\times U_R(1)\rightarrow U_V(1)$ will be studied in presence of an explicit symmetry breaking term in the effective action. This term, however, will be quadratic in the complex scalar field \cite{brauner05,benes07}, in contrast to the commonly assumed linear dependence.

Model systems with an Abelian continuous symmetry have received constant interest in the past. The symmetric part of the Lagrangian has the same symmetry as the chiral one-flavor NJL model, which served as the most transparent illustration of the symmetry breaking mechanism proposed in the classic publication of Nambu and Jona-Lasinio \cite{nambu61}. It has been exploited as the simplest example of dynamical fermion mass generation in discussions of chiral symmetry breaking in hadrons and nuclei at finite temperature and baryonic density \cite{vogl91}. Detailed investigations of its fixed point structure were realized with emphasis on the Fierz ambiguity of the parametrization of its effective action \cite{braun12,braun17}. Its bosonization with help of a Hubbard-Stratonovich (HS) transformation \cite{s58,h58} leads to a Yukawa-type model rather similar to the one investigated in this paper. The complex scalar field is of an auxiliary nature, it stands for a bound fermion-anti-fermion state. It has no proper dynamics at the scale, where the HS-transformation is applied, though field fluctuations generate  in the infrared a kinetic term and also higher powers of the field in its potential energy. 

It can be noted that the compositeness scale defined in \cite{bardeen90,hasenfratz91} through the vanishing of the wave function renormalization constant of the  scalar field could be included also into the present discussion by applying existing renormalization group techniques \cite{jungnickel96,gies02}.
The assumed large amplitude evolution of the field renormalization (the anomalous dimension) of the scalar field due to the strong Yukawa interaction between the scalar and the fermion field was the idea behind the suggestion of the composite ($t-\bar t$ bound state) nature of the Higgs field. Top-quark condensation was one of the first suggestions \cite{miransky89a,miransky89b,bardeen90, hasenfratz91} for the mechanism of the electroweak symmetry breaking.
  
Another potential physical application of the $U(1)$-symmetric Yukawa model is the Frogatt-Nielsen mechanism explaining the hierarchy of lepton masses \cite{frogatt79}. The starting assumption is that different lepton flavors have different charges under a hypothetical $U(1)$ symmetry, therefore couple to different powers of the bosonic flavon field in an effective low-energy model. The mass ratios can be satisfactorily reproduced by choosing appropriate powers. 

The aim of the present study is to take into account the dynamical effect of quantum fluctuations on all couplings with help of renomalisation group equations (RGE) \cite{wetterich91,wetterich93,morris94}. For a first exploration of the emerging infrared physics the equations will be solved in the local potential approximation (LPA) and the scalar potential is truncated at quartic power. The (in)sensitivity of the results to these restrictions will be discussed at length  when higher powers of the potential and the effects of the wave function renormalization are switched on (the so-called $LPA^\prime$).

The initial (ultraviolet) strength of the explicit symmetry breaking relative to the symmetric mass term of the scalar field will be varied. This variation is hardly noticeable on the location of the critical surface separating the symmetric and broken symmetry phases of the model. At the surface, an Ising-type phase transition occurs corresponding to a $Z(2)$-like symmetry left by the quadratic explicit symmetry breaking term. In the symmetry broken phase the infrared spectrum will be fixed to values motivated by the known spectra of the top-Higgs sector in such a way that the pseudo-Goldstone boson of the broken $U_A(1)$ symmetry is associated with the Higgs particle. The ultraviolet strength of the explicit symmetry breaking is systematically diminished keeping the fermion-to-pseudo-Goldstone mass ratio fixed. One finds that the heavy-boson-to-pseudo-Goldstone mass ratio is governed by the quartic self-coupling and it approaches a unique value in the infrared irrespective of its starting ultraviolet value. The limiting value is determined  by the occurrence in LPA of a line of ultraviolet fixed points in the coupling space. The fully massive fermion-boson spectrum emerging in the symmetric limiting case apparently does not follow Goldstone's theorem. 

There is an obvious interest in investigating this new class of theories possessing in the leading order of the gradient expansion an interacting UV fixed line for the ultraviolet completion of quantum field theories \cite{weinberg79}. A rather general perturbative analysis of the conditions for the existence of a UV fixed point was given recently in Ref. \cite{litim14}. It was followed by a general discussion of the stability of the ground state around a UV fixed point \cite{litim16} and the way symmetry breaking can be induced radiatively \cite{abel17}. Existence of an interacting UV-fixed point in a Yukawa-coupled system of Goldstone bosons and chiral fermions was discussed perturbatively in Ref.\cite{bazzocchi11}.  Necessary and sufficient conditions for asymptotic safety in general weakly coupled gauged theories were constructed \cite{bond16} and collider based tests of some asymptotically safe extensions of the Standard Model were investigated \cite{bond17}. Nonperturbative exploration has been achieved for a number of field theoretical models, including Yukawa-models of chiral fermions and bosons with the help of functional renormalization group equations \cite{gies09,gies10,braun11,gies13,jakovac15}. It should be emphasized that the quadratic explicit symmetry breaking plays a role possibly controlling the UV quantum fluctuations even in the exact solutions of these systems.
   
 One notes that this model with dynamically stabilized  quadratic explicit symmetry breaking parameters has been discussed in the symmetric phase with the help of Dyson-Schwinger equations with all symmetry preserving couplings held fixed \cite{brauner05,benes07}.

The paper is organized as follows. In Sec. \ref{model-rge} the effective quantum action investigated in this paper is introduced and the RGE's of its couplings are presented in $LPA^\prime$ both in
the symmetric and the broken symmetry phase based on the general Wetterich equation, supplemented with a linear cutoff function \cite{litim01}. In Sec. \ref{phase-structure} the phase structure is mapped out in $LPA$ for different initial strengths of the explicit quadratic symmetry breaking term  with a quartic scalar potential. It is shown that a critical Yukawa coupling value separates the phases of broken and unbroken $U_A(1)$ symmetry for each initial set of the couplings characterizing the potential energy of the scalar field. This value is determined first from solving the RGE's with scale independent Yukawa coupling. Next, it is argued that both the effect of the Yukawa-running and of the initial strength of the explicit symmetry breaking on the location of the critical surface is negligible. In Sec. \ref{lpa-spectra} the curves of constant pseudo-Goldstone-to-fermion mass ratios are found in the coupling space and the accessible range of the heavy-boson-to-pseudo-Goldstone mass ratio is determined.  The analysis of the mass spectra is systematically repeated upon diminishing the relative strength of the quadratic explicit symmetry breaking. The infrared value of the heavy-to-pseudo-Goldstone mass ratio appears to approach a unique limiting value. The emergence of this limiting solution is interpreted in Sec. \ref{lpa-rgflow} with the help of a line of fixed points arising in $LPA$ in the symmetric phase of the theory. The stability of the fixed point structure is confirmed when higher dimensional operators with moderate strength are included into the scalar potential. In Sec. \ref{wfr-effect} the ansatz for the effective action is extended to include the effects of wave function renormalization ($LPA^\prime$). Although the slow (logarithmic) evolution of the Yukawa coupling effaces the interacting fixed line, but remarkably only an "adiabatic drift" closely following the change in the $LPA$ RG-flow pattern with the variation of the Yukawa coupling is produced. The $LPA$ fixed line still attracts the RG trajectories from the UV-region again leading to a narrowing of the range of  mass ratios of the scalar masses when the strength of the explicit symmetry breaking gets weaker.  A summary of our results accompanied with an outline of its possible extension to more general field theoretical models is given in Sec. \ref{conclusion}. An appendix provides detailed explicit information on the renormalisation group equations (RGE's) used in this study.

\section{The model and its RGE's}
\label{model-rge}

The invariant part of the action is the following:
\bea
\Gamma_{INV}&=&\int d^dx\left[Z_\psi\bar\psi\sls{\partial}\psi+Z_\phi\partial_m\Phi^*\partial_m\Phi+U(\Phi^*\Phi)+h(\bar\psi_R\psi_L\Phi^*+\bar\psi_L\psi_R\Phi)\right],\nonumber\\
 \Phi&=&\frac{1}{\sqrt{2}}(\Phi_1(x)+i\Phi_2(x)),\qquad \psi_{R/L}=\frac{1}{2}(1\pm\gamma_5)\psi.
\label{sym-action}
\eea
For the potential one writes in the symmetric phase
\be
U^{(INV)}=M^2\Phi^*\Phi+\frac{\lambda}{6}(\Phi^*\Phi)^2+\lambda_3(\Phi^*\Phi)^3,
\label{pert-U-SYM}
\ee
while in presence of a condensate the parametrization
\be
U^{(SSB)}=\frac{\lambda}{6}\left(\Phi^*\Phi-\frac{v^2}{2}\right)^2+\lambda_3\left(\Phi^*\Phi-\frac{v^2}{2}\right)^3
\label{pert-U-SSB}
\ee
is more convenient. The $\sim(\Phi^*\Phi)^3$ piece, which completes the perturbatively renormalizable part of the potential, will be used in Sec. \ref{lpa-rgflow} as a small perturbation for testing  the robustness of the findings obtained with quartic potential. In other sections we set $\lambda_3=0$. No attempt will be made to explore the region $\lambda<0,~\lambda_3>0$.

The action (\ref{sym-action}) has two global $U(1)$ symmetries. The first corresponds to the fermion number conservation,
\be
\psi\rightarrow e^{i\alpha}\psi,\qquad \Phi\rightarrow \Phi,
\ee
the second is generated by axial $U(1)$ transformations,
\be
\psi\rightarrow e^{i\gamma_5\Theta}\psi,\qquad \bar\psi\rightarrow \bar\psi e^{i\gamma_5\Theta},\qquad \Phi\rightarrow e^{-2i\Theta}\Phi,
\ee
which is written for the fermions of definite chiral projection as
\be
\psi_L\rightarrow e^{-i\Theta}\psi_L,\qquad \psi_R\rightarrow e^{i\Theta}\psi_R.
\ee
This last symmetry does not allow the presence of a chirality changing mass term for the fermions and also requires vanishing of $\Gamma^{(2)}_{\Phi\Phi}=\delta^2\Gamma/\delta\Phi^2$ and $\Gamma^{(2)}_{\Phi^*\Phi^*}=\delta^2\Gamma/\delta\Phi^{*2}$ in the effective action. Dynamical emergence of the corresponding terms in the course of the RG evolution corresponds to dynamical symmetry breaking (DSB) of the axial $U_A(1)$ symmetry,
\be
\Gamma_{DSB}=\int_p[\Pi\Phi(-p)\Phi(p)+\Pi^*\Phi^*(-p)\Phi^*(p)]+
\int_p[\Sigma_{LR}\bar\psi_L(-p)\psi_R(p)+\Sigma_{RL}\bar\psi_R(-p)\psi_L(p)].
\label{breaking-action}
\ee
In the present investigation we set $\Sigma_{RL}=\Sigma_{LR}=0$, which means that the fermion mass is fully due to the spontaneous breaking of the $U_A(1)$ symmetry. Then there is a $Z(2)$-like remaining discrete symmetry of the system,
\be
\Phi\rightarrow -\Phi,\qquad \psi_L\rightarrow i\psi_L, \qquad \psi_R\rightarrow -i\psi_R.
\label{ising-sym}
\ee
\subsection{RGE's of the system without scalar condensate}

The Wetterich equation for a fermion-boson system can be partitioned in the following form \cite{jakovac13,aoki97}:
\bea
\partial_t\Gamma=\frac{1}{2}\hat\partial_t{\textrm{STr}}\log(\Gamma^{(2)}+R_k)
&=&-\frac{1}{2}\hat\partial_t{\textrm{Tr}}\log(\Gamma^{(2)}_{F}+R_k^F)
+\frac{1}{2}\hat\partial_t{\textrm{Tr}}\log(\Gamma^{(2)}_{B}+R_k^B)\nonumber\\
&+&\frac{1}{2}\hat\partial_t{\textrm{Tr}}\log\left[1-(\Gamma^{(2)}_B+R_k^B)^{-1}\Gamma^{(2)}_{BF}(\Gamma^{(2)}_F+R_k^F)^{-1}\Gamma_{FB}^{(2)}\right].
\label{wetterich-eq}
\eea
Here $\Gamma^{(2)}$ denotes the full second functional derivative of the effective action, while $\Gamma^{(2)}_B$ and $\Gamma^{(2)}_F$ refer to its purely bosonic and purely fermionic sector, respectively. $\Gamma^{(2)}_{FB}$ and $\Gamma^{(2)}_{BF}$ each contain one fermionic and one bosonic functional derivative. The last term is present when the system is immersed in a nonzero fermionic background as was discussed at length in Refs.\cite{jakovac13,aoki97}. It contributes to specific $n$-point functions defined through a number of fermionic functional derivatives even if the fermionic condensate is not present. The cutoff functions $R_k^{B/F}$ restrict the functional trace to field components with momenta larger than the actual scale $k$.
The operation $\hat\partial_t=k\hat\partial_k$ acts only on the scale dependence of the cutoff functions $R_k^{B/F}$. 

The bosonic part of $\Gamma^{(2)}$ can be written with $U$ given in (\ref{pert-U-SYM}) in the following matrix form when one uses the row vector $(\Phi^*,\Phi)$ and 
its adjungated column vector as independent variables:
\be
\displaystyle
\Gamma_B^{(2)}(q,q')=\frac{\delta(q+q')}{(2\pi)^d}
\begin{pmatrix}
Z_\phi q^2_R+M^2 & 2\Pi^*\\
2\Pi&Z_\phi q^2_R+M^2
\end{pmatrix},
\label{inverse-bosonic}
\ee
where $q^2_R=q^2+R^B(q)$.
It gives the propagator 
\be 
G_B(q,q')=\frac{(2\pi)^d\delta(q+q')}{\Delta(q^2_R)}
\begin{pmatrix}
Z_\phi q^2_R+M^2 & -2\Pi\\
-2\Pi^*&Z_\phi q^2_R+M^2
\end{pmatrix}.
\ee
The zeros of the determinant of $\Gamma^{(2)}_B$ determine the bosonic spectra at the actual scale:
\be
\Delta(q^2_R)=(Z_\phi q^2_R+M^2)^2-4|\Pi|^2,\qquad m_{hb}^2=M^2-2|\Pi|,\qquad m_G^2=M^2+2|\Pi|.
\ee
The meaning of the indices "hb" and "G" will be clarified when discussing the spontaneously broken phase. In the symmetric phase the fermion has a massless propagator and no mass term is generated in this phase by the quantum fluctuations either.

In the symmetric phase the RGE's of the parameters characterizing the quartic potential and the Yukawa interaction are the following:
\begin{eqnarray}
&&
\partial_tM^2=-2h^2\int_q\hat\partial_t\frac{1}{Z_\psi^2 q_R^2}+\frac{2\lambda}{3}\int_q\hat\partial_t\frac{Z_\phi q_R^2+M^2}{\Delta(q_R^2)},\nonumber\\
&&
\partial_t\lambda=6h^4\int_q\hat\partial_t\frac{1}{Z_\psi^4q_R^4}-\frac{\lambda^2}{3}\int_q\hat\partial_t\frac{5(Z_\phi q_R^2+M^2)^2+16|\Pi|^2}{\Delta^2(q_R^2)},\nonumber\\
&&
\partial_th=h^3\Pi\int_q\hat\partial_t
\frac{1}{Z_\psi^2q_R^2\Delta(q_R^2)}.
\label{RGE-SYM}
\end{eqnarray}
The evolution of the explicit quadratic symmetry breaking parameter is driven by itself:
\be
\partial_t\Pi=-\frac{\lambda\Pi}{3}\int_q\hat\partial_t\frac{1}{\Delta(q_R^2)}.
\label{RGE-PI}
\ee

The condition for the existence of the symmetric phase with vanishing scalar background $\Phi_b=\Phi^*_b=u/\sqrt{2}$ can be read off from the one-point equation:
\be
\frac{\delta\Gamma}{\delta\Phi}\Bigr|_{\Phi_b}=U'\Phi_b^*+2\Pi\Phi_b=
\left(M^2+2\Pi+\frac{\lambda}{6}u^2\right)\frac{u}{\sqrt{2}}=0.
\ee
The phase transition to the broken symmetry phase occurs where $M^2+2\Pi$ vanishes. For this reason it is convenient to  choose $\Pi<0$. At the phase transition one has to change the parametrization of the potential to (\ref{pert-U-SSB}).

The $\hat\partial_t$ operation has been applied to the integrands of (\ref{RGE-SYM}) and (\ref{RGE-PI}) with the choice of the linear (or optimized) cutoff functions \cite{litim01}.
Then introducing the rescaled variables
\bea
&\displaystyle
\Pi_r=\frac{\Pi}{Z_\phi k^2},\qquad
h_r^2=Z_\psi^{-2}Z_\phi^{-1}h^2k^{d-4},\qquad M_r^2=\frac{M^2}{Z_\phi k^2},\qquad
\lambda_r=Z_\phi^{-2}\lambda k^{d-4},\nonumber\\
&\displaystyle
u_r=Z_\phi^{1/2}uk^{(2-d)/2},\qquad v_r=Z_\phi^{1/2}vk^{(2-d)/2},
\eea
one finds the explicit expressions for the RGE's in terms of dimensionless quantities in $LPA^\prime$. The complete set, including the RG running of the anomalous dimensions $\eta_i$,
\be
\eta_\psi=-\partial_t\ln Z_\psi,\qquad \eta_\phi=-\partial_t Z_\phi
\ee
 is given in the Appendix.

\subsection{RGE's of the system with scalar condensate}

The equation for the condensate takes now the form,
\be
\left(\frac{\lambda}{6}(u^2-v^2)+2\Pi\right)\frac{u}{\sqrt{2}}=0.
\label{condensate-eq-SSB}
\ee
It has a positive solution for $u^2$ under the condition $\lambda v^2-12\Pi>0$. This inequality is automatically fulfilled for negative values of $\Pi$. Below everywhere $|\Pi|$ will be displayed.

The particle spectra is determined by the following expressions:
\bea
&\displaystyle
\Delta(q^2_R)=\left(Z_\phi q_R^2+\frac{\lambda}{3}v^2+4|\Pi|\right)(Z_\phi q_R^2+4|\Pi|),\nonumber\\
&\displaystyle
m_{hb}^2=\frac{\lambda}{3}v^2+4|\Pi|=\frac{\lambda}{3}u^2,\qquad m_G^2=4|\Pi|,\qquad m_\psi=h\frac{u}{\sqrt 2}.
\label{SB-spectra}
\eea
The pseudo-Goldstone mass is given by $m_G$, the mass of the heavy boson is denoted by $m_{hb}$.

In the broken symmetry phase the parameter $v^2$ replaces $M^2$. In this phase it is more convenient to solve the RG equations of the masses, where one exploits the relation  $\lambda v^2/3=m_{hb}^2-m_G^2$ which follows (\ref{condensate-eq-SSB}) and (\ref{SB-spectra}). The RGE of $|\Pi|$ differs from that of $m_G^2$ only by a trivial $1/4$ factor. One writes the following two equations:
\begin{eqnarray}
&&
\frac{3}{4}\partial_tm_G^2=-2h^2\hat\partial_t\int_q\frac{1}{Z_\psi^2q_R^2+m_\psi^2}+\frac{\lambda}{6}\hat\partial_t\int_q\frac{4Z_\phi q_R^2+m_{hb}^2+m_G^2}{\Delta(q_R^2)},\nonumber\\
&&
\partial_tm_{hb}^2=2h^2\hat\partial_t\int_q\frac{-Z_\psi^2q^2_R+m_\psi^2}{(Z_\psi^2q_R^2+m_\psi^2)^2}\\
&&
+\frac{\lambda}{6}\hat\partial_t\int_q\frac{1}{\Delta(q_R^2)}\left[4Z_\phi q_R^2+7m_{hb}^2+3m_G^2-\frac{m_{hb}^2}{\Delta(q_R^2)}\left(4Z_\phi q_R^2+m_{hb}^2+m_G^2\right)^2\right].\nonumber
\label{RGE-SB-1}
\end{eqnarray} 

In order to have sufficient number of equations one makes use of the equation of $h$ and $\lambda$, expressing their rates through the masses. The RGE for the Yukawa coupling is the following:
\begin{eqnarray}
\partial_th&=&-\frac{h^3}{2}(m_{hb}^2-m_G^2)\hat\partial_t\int_q
\frac{-Z_\psi^2q_R^2+m_\psi^2}{(Z_\psi^2q_R^2+m_\psi^2)^2\Delta(q_R^2)}\nonumber\\
&+&\frac{h^3m_{hb}^2}{4}\hat\partial_t\int_q\frac{1}{Z_\psi^2q_R^2+m_\psi^2}\left[\frac{3}{(Z_\phi q_R^2+m_{hb}^2)^2}-\frac{1}{(Z_\phi q_R^2+m_G^2)^2}\right].
\label{RGE-SB-2}
\end{eqnarray}
The second term on the right-hand side of the equation is the result of the $\Phi$ dependence of the boson propagator and is present only in the broken symmetry phase.
 For the quartic coupling the simplest is to use its original definition: $(3/2)\delta^4\Gamma/(\delta\Phi^2\delta\Phi^{*2})$. One finds a rather lengthy expression due to the complicated dependence of the determinant $\Delta$ on $\Phi$ and $\Phi^*$. Therefore we give its explicit expression only in the Appendix, together with
 the $LPA^\prime$ equations for the scaled (dimensionless) coupling strengths.

\section{Phase structure in $LPA$}
\label{phase-structure}

The phase structure of the system in $d=4$ has been mapped out in the three-dimensional coupling space of the microscopic parameters $(\lambda_{\Lambda},M^2_{r\Lambda},h_{\Lambda})$ characterizing the system at the UV scale $\Lambda$.
RG evolutions starting from the symmetric phase were followed to $k=0$, since it is natural to assume that the "microscopic" theory is in the symmetric phase. The RG trajectory of the system was traced for various initial ratios $|\Pi_\Lambda|/M^2_\Lambda$, and our goal was to see to what extent the resulting physical picture depends on this ratio. In the exploratory RG runs the Yukawa coupling $h$ was kept constant, which was a fortunate choice, since this turned out to be increasingly good approximation with diminishing the ratio $|\Pi_\Lambda|/M^2_\Lambda$. In fact, in the symmetric phase the rate of variation of $h$ is proportional to $\Pi(k)$ [see Eq.(\ref{RGE-SYM})]. The Yukawa coupling turned out to control the borderline between the symmetric 
and the broken $U_A(1)$ symmetry phase of the model.

\begin{figure}
\begin{center}
\includegraphics[width=6cm]{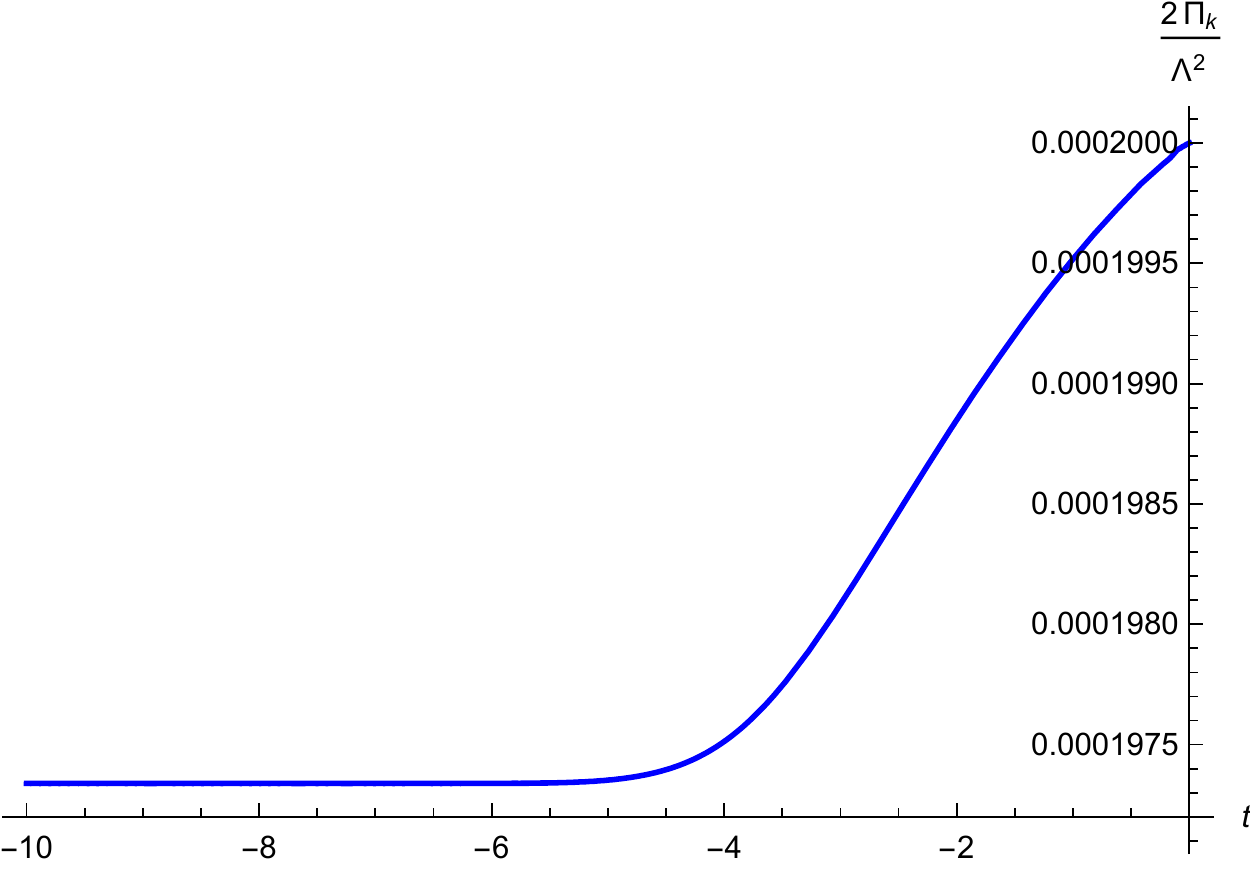}
\includegraphics[width=6cm]{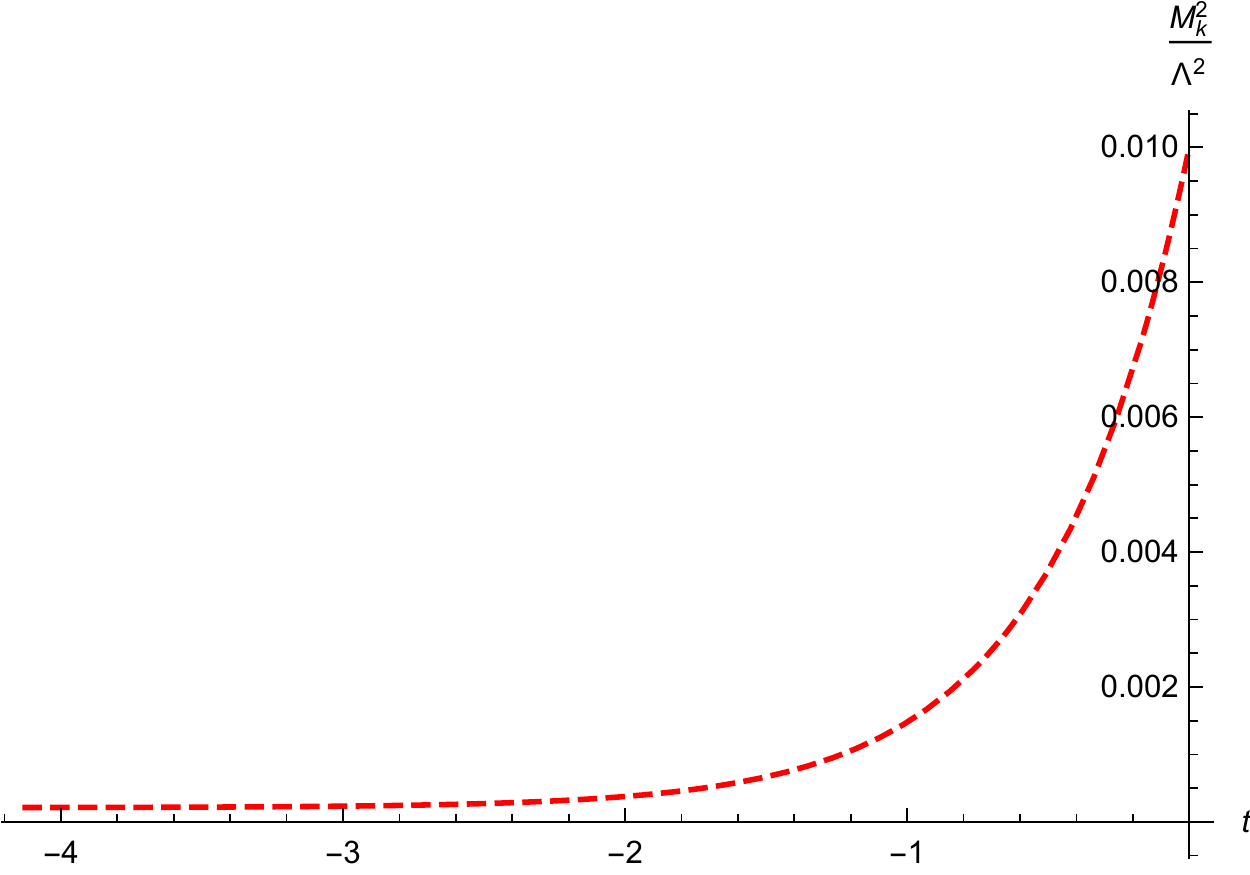}
\end{center}
\caption{A typical RG evolution of $2|\Pi|$ (left) and $M^2$ (right) measured in units of $\Lambda^2$ for $|\Pi_\Lambda|/M_\Lambda^2=0.01$}
\label{pi-running-m2-running}
\end{figure}

\begin{figure}
\begin{center}
\includegraphics[width=9cm]{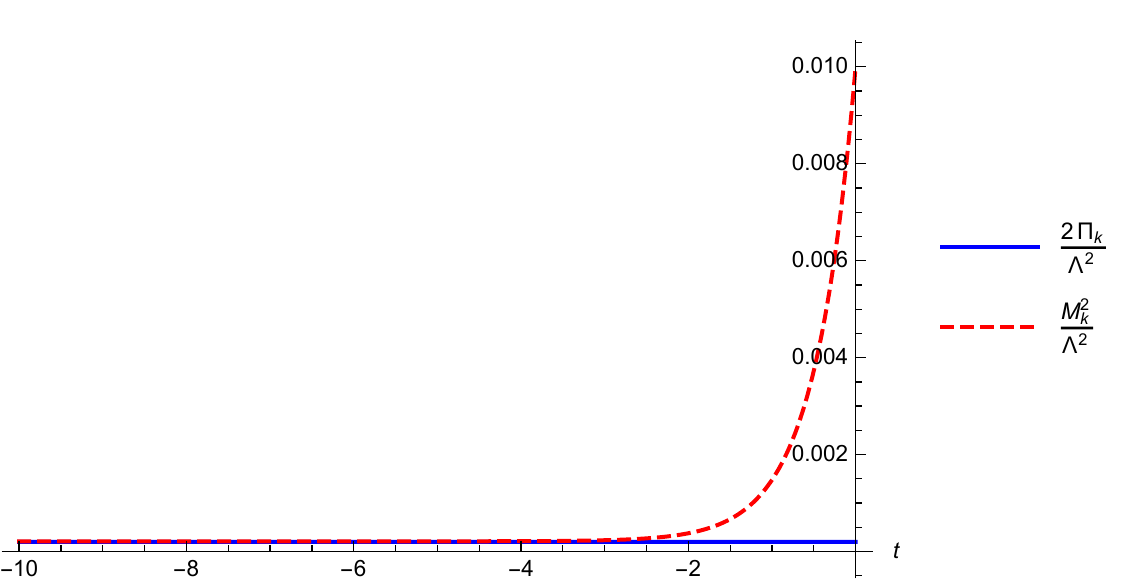}
\end{center}
\caption{An RG evolution of $M^2$ and $2|\Pi|$ ending on the critical surface for $|\Pi_\Lambda|/M_\Lambda^2=0.01$}
\label{overlay-pi-running-m2-running}
\end{figure}

A typical run is displayed in Fig.\ref{pi-running-m2-running}, for the starting ratio $|\Pi_\Lambda|/M_\Lambda^2=0.01$.
The initial data of this run is chosen to end up at $k=0$ with $M_0^2/\Lambda^2=2|\Pi_0|/\Lambda^2$, e.g., in a point on the critical surface separating the symmetric and the spontaneously broken phases as one sees when overlaying the two curves in Fig.\ref{overlay-pi-running-m2-running}. Although the small value of $|\Pi_\Lambda|/\Lambda^2$ drops by circa 2\%, it is stabilized quickly and actually attracts the value of $M^2_k/\Lambda^2$ starting much higher and steeply decreasing by the term corresponding to the quadratic divergence of the conventional perturbation theory. This qualitative behavior is universal in the sense that it is independent of the initial $|\Pi_\Lambda|/M^2_\Lambda$ ratio. 

\begin{figure}
\begin{center}
\includegraphics[width=9cm]{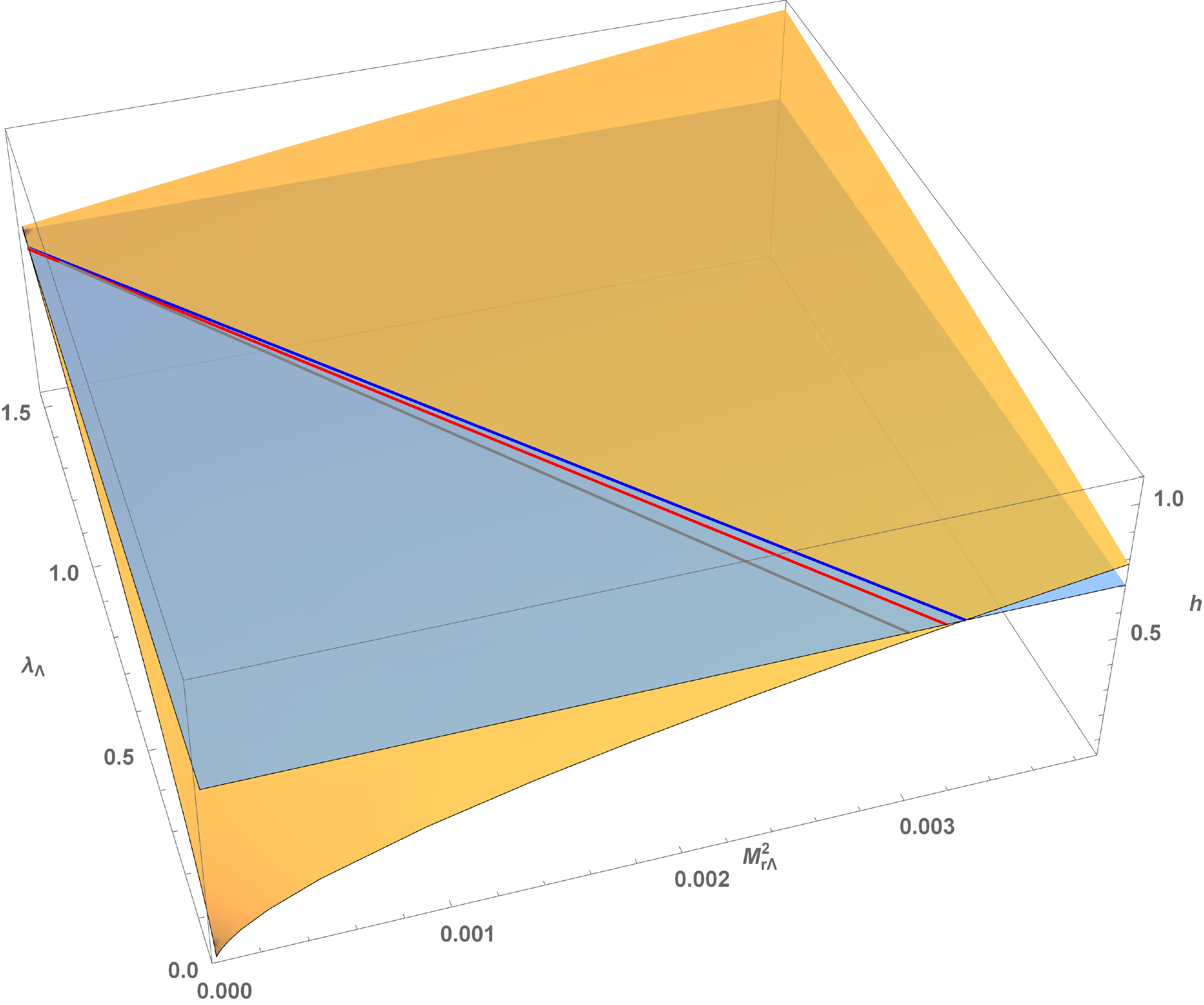}
\end{center}
\caption{The critical surface determined for the ratio $|\Pi_\Lambda|/M^2_\Lambda=0.01$ (yellow). The region belonging to the broken symmetry phase lies above the critical surface, the plane corresponding to the fixed value of the Yukawa coupling is colored blue. Further decreasing the value of the relative strength of the explicit symmetry breaking causes essentially no change in the shape of the critical surface. Also lines of equal infrared values of the $m_G^2/m_\psi^2$ ratio are displayed for two different values.}
\label{critical-surface-cut-h}
\end{figure}
It turns out that for every initial ($M_{r\Lambda}^2, \lambda_\Lambda$) pair the trajectory stays in the symmetric phase when one chooses $0<h_\Lambda<h_c(M_{r\Lambda}^2,\lambda_\Lambda)$. The critical surface $h_c(M_{r\Lambda}^2,\lambda_\Lambda)$ is displayed in Fig.\ref{critical-surface-cut-h}, where the surface is cut by the "physical" value of $h_{phys}=173/246$, corresponding to the toy Higgs-top model. 
Since the physical world realizes the spontaneously broken phase, the allowed region of the parameter space is nearly triangular, where $h_{phys}>h_c(M_{r\Lambda}^2,\lambda_\Lambda)$.

The variation of the surface (and also of the edge of the region belonging to the broken symmetry phase) observed for larger values of $|\Pi_\Lambda|/M^2_\Lambda$, becomes unobservably small when $|\Pi_\Lambda|/M_\Lambda^2\ll 0.01$ is chosen. The critical surface converges to a limiting shape essentially coinciding with the surface displayed in the figure. Also it is important to note that the effect of the RG running of $h$ in the final values of either $M^2(k=0)/\Lambda^2$ or $2|\Pi(k=0)|/\Lambda^2$ shows up only in the fifth decimal place.

The universality class of the phase transition can be characterized by the power of the dependence of the difference $M^2(k=0)/2|\Pi(k=0)-1|$ on $h-h_c$, as approached from the symmetric phase. Since the lower mass in this phase has a squared mass $M^2-2|\Pi|$, this difference describes the infinite increase of the correlation length (the inverse mass) when the critical surface is approached. Its scaling was tested in several points of the critical surface and found to be linear, which corresponds to a critical exponent $\nu\approx 1/2$, coinciding with the mean-field value of the correlation length exponent of the Ising model. This conclusion is in agreement with the fact that the quadratic explicit symmetry breaking leaves for the full system a $Z(2)$-like symmetry intact [see Eq.(\ref{ising-sym})]. 

The order parameter and the excitation spectra in the broken symmetry phase, however, reflects the $U_A(1)$ symmetry. In particular, one finds a pseudo-Goldstone excitation in this phase. It is rather interesting that a  sort of level crossing phenomenon is produced through the critical surface. The pseudo-Goldstone mode is continuously connected with the more massive mode of the symmetric phase, while the heavy mode of the broken symmetry phase develops from the critical excitation.

\section{Spectra of the broken symmetry phase in $LPA$}
\label{lpa-spectra}

As a first step of the exploration of the spectra one can locate the starting $(\lambda_\Lambda, M^2_{r\Lambda})$ points wherefrom 
a fixed infrared $m_G^2/m_\psi^2$ ratio can be reached. In the allowed region ($h>h_c$), 
which has a triangular shape when looking at it almost parallel to the $h$ axis, these points form an almost linear curve in the $\lambda_\Lambda-M^2_{r\Lambda}$-plane as one can see from
Fig.\ref{critical-surface-cut-h}. This figure is based on the RG evolutions done with $|\Pi_\Lambda|/M^2_\Lambda=0.01$, but its variation had not affected the qualitative features of the fixed mass-ratio curves. It is remarkable that the red-colored line which corresponds to the ratio when the pseudo-Goldstone is identified with the Higgs boson, and the fermion with the mass of the top, passes extremely close to criticality. This circumstance explains that even in the broken symmetry phase, the Yukawa coupling is hardly changing. In this region the additional term $\sim m_{hb}^2$ of its rate of evolution is also rather small, see Eq.(\ref{RGE-SB-2}).

\begin{figure}
\begin{center}
\includegraphics[width=11cm]{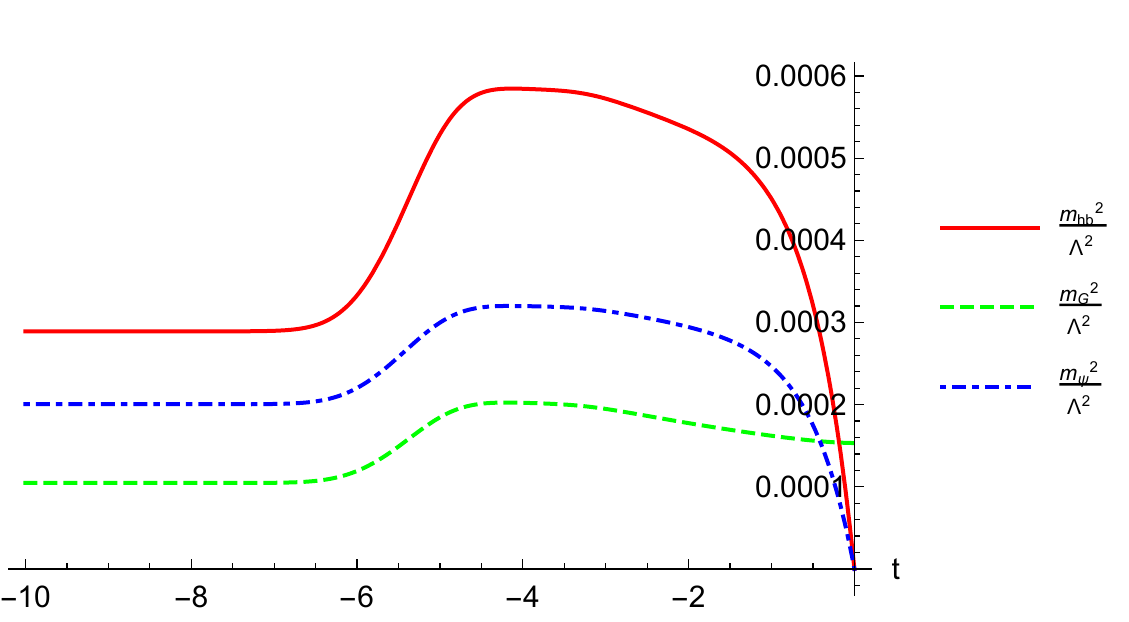}
\end{center}
\caption{RG evolution of the dimensionless squared fermion and boson masses (the PNGB is the lightest at  $t=-\infty$). Note that the evolutions of the masses are displayed only in the broken symmetry phase (therefore $m^2_{hb}$ and $m_\psi^2$ start from zero).}
\label{h-H-evolution}
\end{figure}

Next, one can investigate the ratio $m_{hb}^2/m_G^2$ along this line in order to see the existence of any bound for the heavy mass. The separate evolution of the two scalar masses in the broken symmetry phase displays a rather nontrivial evolution. Also the fermion ("top"-) mass displays a sort of "walking" evolution around the same value of $t=\ln(k/\Lambda)$ as can be seen from Fig.\ref{h-H-evolution}.

\begin{figure}
\begin{center}
\includegraphics[width=9cm]{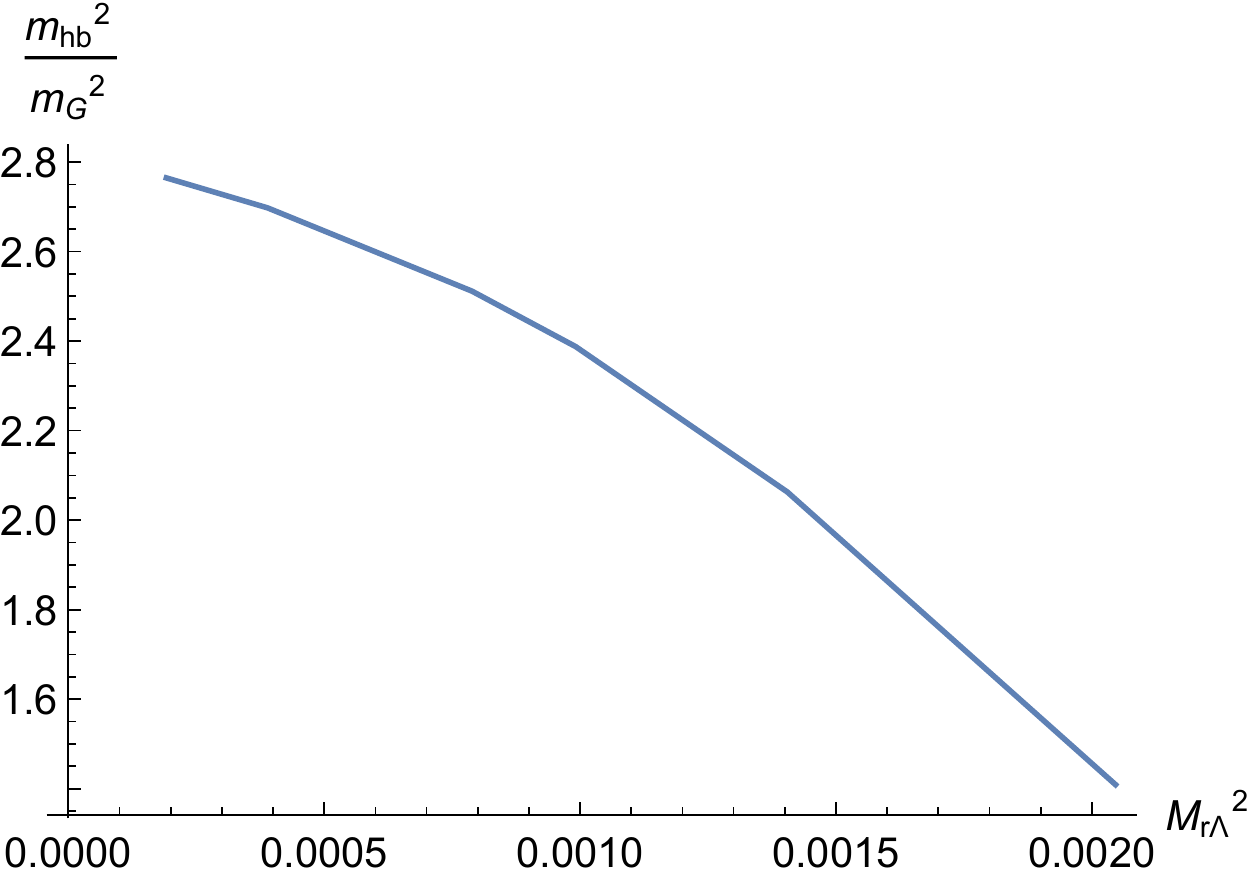}
\end{center}
\caption{Monotonic variation of the heavy-light scalar mass ratio with $M_{r\Lambda}^2$ for $|\Pi_\Lambda|/M^2_\Lambda=.01$}.
\label{mh2/mH2-ratio}
\end{figure}

The ratio $m_{hb}^2/m_G^2$ shows monotonic increase with decreasing $M_{r\Lambda}^2$ (or equivalently, increasing $\lambda_\Lambda$). Figure \ref{mh2/mH2-ratio} clearly shows that there is an upper bound at the present starting $|\Pi_\Lambda|/M^2_\Lambda$ ratio.

The interesting question is to see how this upper bound modifies when the strength of the explicit symmetry breaking is systematically diminished. Since the form of the effective action is unchanged one has for the mass ratios (see Eq.(\ref{SB-spectra})),
\be
\frac{m_G^2}{m_\psi^2}=\frac{8|\Pi(k=0)|}{h^2u^2(k=0)},\qquad  \frac{m_G^2}{m_{hb}^2}=\frac{12|\Pi(k=0)|}{\lambda(k=0) u^2(k=0)}=\frac{3h^2(k=0)}{2\lambda(k=0)}\frac{m_G^2}{m_\psi^2}.
\label{scalar-ratio}
\ee
The first ratio and $m_\psi/u/\sqrt{2}$ were fixed to prescribed values. Then the variation of the second simply reflects the variation of $\lambda(k=0)$ with the initial data. The obvious question is to see the dependence of $\lambda(k=0)$ on $\Pi_\Lambda/M^2_\Lambda$ when its evolution starts with an arbitrary $\lambda_\Lambda$.

\begin{figure}
\begin{center}
\includegraphics[width=11cm]{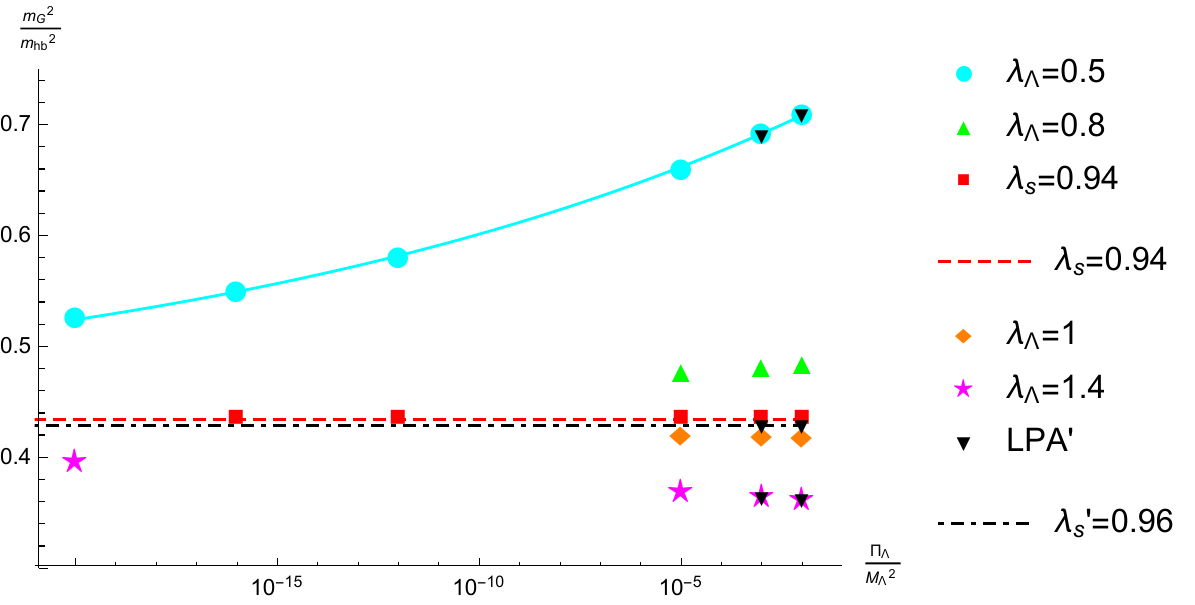}
\end{center}
\caption{Variation of $m_G^2/m_{hb}^2$ in $LPA$ with $\Pi_\Lambda/M^2_\Lambda$ for various $\lambda_\Lambda$ values ($h_\Lambda=173/246$). Stability of results obtained in $LPA$ is checked with $LPA^\prime$ for parameter sets, indicated in the figure (black triangles  and dash-dotted black horizontal line).}
\label{explicit-SB-dependence}
\end{figure}

In Fig.\ref{explicit-SB-dependence}, this variation is displayed in $LPA$ for $\lambda_\Lambda=0.5,0.8,0.94,1.0,1.4$ (this figure contains information also concerning the behavior in $LPA^\prime$, which will be discussed in Sec. VI).
The first thing to observe is a definite dependence with diminishing  $\Pi_\Lambda/M^2_\Lambda$ in the range $10^{-3}-10^{-7}$. This is rather different from the case of the linear explicit symmetry breaking [$\sim H(\Phi+\Phi^*)$], where $\lambda$ does not depend on the strength $H$ of the explicit breaking. Actually, there is a specific  $\lambda_\Lambda$ value equal to $\lambda_s=0.9400$, where $|\Pi_\Lambda|/M_\Lambda^2$ does not have any effect on $m_G^2/m_{hb}^2$. It appears on the figure that the mass ratios arising in the region $\lambda_\Lambda>\lambda_s$ are increasing, while those falling into region $\lambda_\Lambda<\lambda_s$ decrease when the ratio $|\Pi_\Lambda|/M_\Lambda^2$ is diminished. The best fit is of the form $a+b(|\Pi_\Lambda|/M_\Lambda^2)^c$. The values $b=0.3103, c=0.0268$ were fitted, while $a$ was fixed to the mass ratio belonging to $\lambda_\Lambda=\lambda_s =0.9400$:
$m_G^2/m_{hb}^2\rightarrow 0.4340$.
These infrared parameters vary with values of $m_G^2/m_\psi^2$ and of $h$ as will be discussed shortly on an other numerical example below.
 \begin{figure}
\begin{center}
\includegraphics[width=11cm]{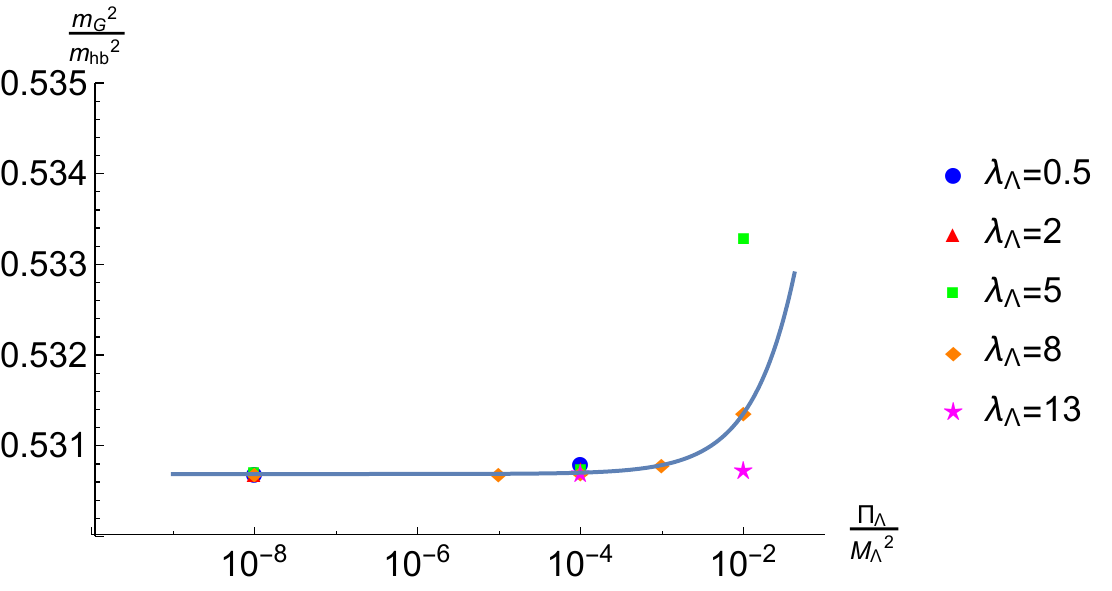}
\end{center}
\caption{Variation of $m_G^2/m_{hb}^2$ in $LPA$ with $\Pi_\Lambda/M^2_\Lambda$ in case of parameters suggested by strong interactions ($h_\Lambda=3$).}
\label{pi-sigma-quark}
\end{figure}

For $\lambda_\Lambda=\lambda_s$ it turns out that all couplings and masses evolve along the same RG trajectory, irrespective the starting $|\Pi_\Lambda|/M^2_\Lambda$ ratio. Just for a smaller starting value the evolution in the symmetric phase becomes longer in $t$. The length of the $t$ interval needed for the buildup of the broken symmetry characteristics is always the same. There is a mapping from the "RG time" $t$ on the values of $|\Pi_\Lambda|/M^2_\Lambda$. This observation is the key for the RG-flow interpretation of the apparent focusing of $\lambda(k=0)$ and with it of the mass ratio $m_{hb}^2/m_G^2$ on a unique finite value, when the quadratic explicit symmetry breaking tends to zero. 

The generic nature of the solution described above is confirmed also when one works out another parametrization suggested by the pion-sigma-constituent quark system of strong interactions. There we chose $u/\sqrt{2}\approx 100$MeV, $m_q\approx 300$MeV and $m_\pi\approx 140$MeV. The convergence to a unique limiting $m_\pi^2/m_\sigma^2$ value is much faster in this case as can be seen in Fig.\ref{pi-sigma-quark}. This example is characterized by the couplings $\lambda_s\approx 17, h=3$, clearly much beyond the perturbative regime. A very good power-law fit is displayed which was determined to the data: $\lambda_\Lambda=8, |\Pi_\Lambda|/M_\Lambda^2=0.01$. The best fit parameters are: $a=0.5307, b=0.0308, c=0.8318$. The value of the exponent $c$ is very different from that obtained in the previous numerical example. Its compatibility with universality is far from obvious.

At this point of the discussion one should retain that in case of the quadratic  explicit symmetry breaking there is an alternative to the conventional picture of spontaneous symmetry breakdown in the limit of vanishing explicit breaking where one keeps the pseudo-Goldstone excitation massive (its ratio to the fermion mass is kept fixed and finite). According to this solution the mass of its heavy boson partner is uniquely predicted in $LPA$. In the next section it is shown that for any choice of the initial $\lambda_\Lambda$ the RG flow unavoidably approaches the renormalized trajectory starting from a UV fixed point $\lambda_{UV}^*$ to which $\lambda_s$ converges when $|\Pi_\Lambda|/M_\Lambda^2\rightarrow 0$.

\section{RG flow in $LPA$ for the limiting case $|\Pi_\Lambda|/M^2_\Lambda\rightarrow 0$}
\label{lpa-rgflow}

It can be very strange, even counterintuitive that, despite we decrease
the strength of the explicit breaking, the system does not approach the spectrum expected in the phases of the fully symmetric
theory. In this section, the reason for
this unexpected behavior is revealed.

Let us start with an analogy that helps us to understand this
phenomenon. In pure Yang-Mills theories the only parameter that
characterizes the system is the coupling constant $g$. It is very
natural to think that a system with large coupling is strongly
interacting, while a small coupling means weak interaction, and in the
zero coupling limit, the system becomes free. Still, this expectation
does not work in Yang-Mills theories. If the coupling is strictly zero,
there is no interaction, of course. But for an arbitrarily
small positive UV value of the coupling, all choices describe the same
physics in the IR limit. The reason is that the coupling $g$ actually depends on
the scale, that is $g=g(k)$, and in the UV limit the coupling goes to
zero $g(k\to\infty)=0$ (asymptotic freedom) - or stated another way,
the system possesses an UV fixed point at $g=0$. There is a one-to-one
correspondence between the value of the coupling and the scale. This
means, however, that giving the value of the coupling only tells us
 the scale, where we are actually looking at the system (this
phenomenon is called dimensional transmutation), and has nothing to do
with the strength of the interaction in the infrared.

Basically the same dimensional transmutation occurs here in our model,
with some important modifications. We have four couplings, $\lambda(k)$, $h(k)$,
$\Pi(k)$ and $M^2(k)$. Their scale dependence in the symmetric phase
is described in LPA by Eq. (\ref{DIMLESS-RGE-SYM}) after putting $\eta_i=0$. Inspecting these
equations, we find that there is in the system an (interacting) UV fixed
point. Assuming $M_r^2\ll1$ (this can be verified {\it posteriori}) we
arrive at the formulas
\begin{equation}
  \frac{\lambda_{UV}^{*2}}{h_{UV}^{4}} \approx \frac{18}5,\qquad 
  \frac{M^{*2}_{r,UV}}{h_{UV}^{2}}\approx 2v_d
  \left(1-\sqrt{\frac{2}{5}}\right),\qquad \Pi_{r,UV}^*=0,
\label{UV-FP-eq}
\end{equation}
while for the value of $h_{UV}$ we do not have any restriction, except that it still should be larger than $h_c(M_{r\Lambda}^2,\lambda_\Lambda)$.
This means that in fact we have a fixed line, not just a point.

Before continuing the analysis of the RG map, we discuss on the example of the operator $\lambda_3\rho^3$  the persistence of the UV fixedline in LPA. The $\lambda_3(\Phi^*\Phi)^3$ term provides a tadpole contribution to the evolution of $\lambda$, proportional to $\lambda_3$. The bosonic contribution to the evolution of $\lambda_3$ consists of two terms as can be checked with (\ref{bosonic-contrib-lambda3}), one is proportional to $\lambda_3\lambda$, the other to $\lambda^3$. There is also a fermionic contribution, Eq.(\ref{fermionic-contrib-lambda3}) proportional to $h^6$. One can solve the fixed point equation of $\lambda_{3r}=\lambda_3k^2$ and substitute its expression into the fixed point equation of $\lambda$. For a first estimate of the effect of this term on the location of $\lambda^*_{UV}$ one can neglect the $M_r^2$ dependence of the new term and substitute into it for $\lambda^*_{UV}$ its crude expression (\ref{UV-FP-eq}). Obviously, the whole contribution will be $\sim h^6$. Comparing this to the $\sim h^4$ term, one instantly derives an upper bound for $h^2$ as negligibility condition for the $\lambda_3\rho^3$ operator. One can look for a numerical solution of the exact fixed point equations for any given value of $h$. For instance, choosing $h=173/246$, one finds $\lambda_{3UV}^*=5.85\times 10^{-4}$ resulting in the following shift $\lambda^*_{UV}: 0.9400\rightarrow 0.9450, M_{rUV}^{*2}: 1.152\cdot 10^{-3}\rightarrow 1.142\times 10^{-3}$. RG trajectories starting from the neighborhood of the new position of the fixed point will quickly find their way back to the trajectories determined with the quartic potential, since $\lambda_{3r}$ diminishes quadratically towards the infrared.
  
\begin{figure}[htbp]
  \centering
  \includegraphics[width=9cm]{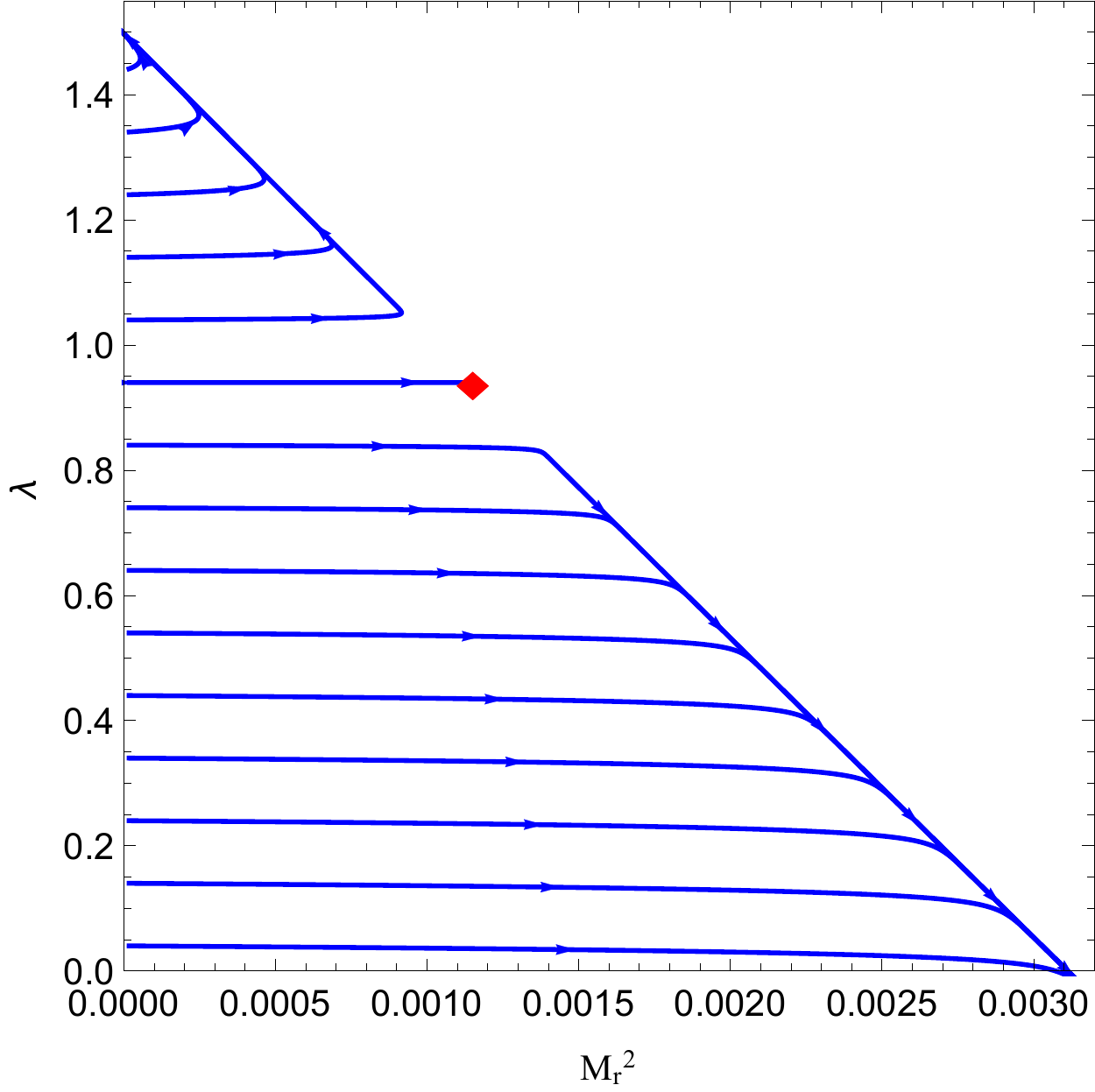}
  \caption{Flow diagram in the $\lambda-M_r^2$ space, restricting to
    those initial values that lead to SSB. We can recover the UV fixed
    point at the analytically determined $\lambda_{UV}^*$ and $M_{rUV}^{*2}$ position (red diamond). 
The arrows point in the direction of increasing the RG time $t$.
 The apparent merging of the RG
    flows along the UV-unstable direction is fake, it is just the consequence of the finite resolution
    of the picture.}
  \label{fig:flow1}
\end{figure}

Let us return to the analysis of the RG map obtained with quartic scalar potential.
In the $\lambda-M_r^2$ space we can map out the flow diagram as shown in
Fig. \ref{fig:flow1}.
In this figure, we restricted ourselves to those UV values that lead to
$SSB$ in the IR. We can recover the UV fixed point at some definite
$\lambda^*$ and $M_r^{*2}$ position, and we see that the fixed point coordinates coincide rather well with the estimate (\ref{UV-FP-eq}) for the chosen value of $h$.  Moreover, using the approximate expression of $\lambda_{UV}^*$ in (\ref{scalar-ratio}), one finds for the scalar mass ratio corresponding to the renormalized trajectory $m_G^2/h_{hb}^2=0.42$, which is in very good agreement with the numerical solution. If we are
on a RG trajectory where $\lambda>\lambda_{UV}^*$, then $\lambda(k)$ will
run into a Landau pole at some $k=\Lambda_L$ scale, where
$M_r^{2}\to -\infty$: this is the upper part of this figure. On the
other hand if we have $\lambda<\lambda_{UV}^*$ at a given scale,
$\lambda(k)$ will cross zero at some $k=\Lambda_0$ value (vacuum
stability bound), where $M_r^2(k)$ remains finite. We remark that
although it seems from this figure that the RG trajectories merge, but
this apparent phenomenon is just due to the finite resolution: two RG trajectories must
never merge or cross each other.

The RG flow of $\Pi_r$ is very dull: it goes to $\Pi_r=0$ with an exponent
very close to $-2$, i.e., $\Pi_r(k)= C/k^2$ for large $k$ values. But this
also means that near the UV fixed point $\Pi/M^2$ scales as
$\sim 1/k^2$. So, similarly to the QCD case, the 
quantity $\Pi/M^2$ has nothing to do with the strength of the
explicit breaking, but tells us the scale, at which we look at the
system.

In the above limiting process towards the vanishing of the explicit symmetry breaking we keep $\lambda_{\Lambda}$ constant,
while decreasing the value of $|\Pi_\Lambda|/M_\Lambda^2$. As we have seen,
$|\Pi|/M^2\to0$ implies increasing scale $k\to\infty$, while we keep
the coupling $\lambda_\Lambda$ fixed upon this process. But, as we also analyzed, at large
scales a generic $\lambda$ blows up or crosses zero, so the fact alone
that at a large scale $k$ it is forced to assume a finite positive value, means that the
RG trajectory on which it actually lies is closer and closer to the renormalized trajectory. Therefore, the corresponding IR value $\lambda(k=0)$ will converge to the one evolving along the renormalized trajectory from the fixed point value $\lambda_{UV}^*$ and to the mass ratio determined by $\lambda_{UV}^*/h_{UV}^2$.

\begin{figure}[htbp]
  \centering
  \includegraphics[width=9cm]{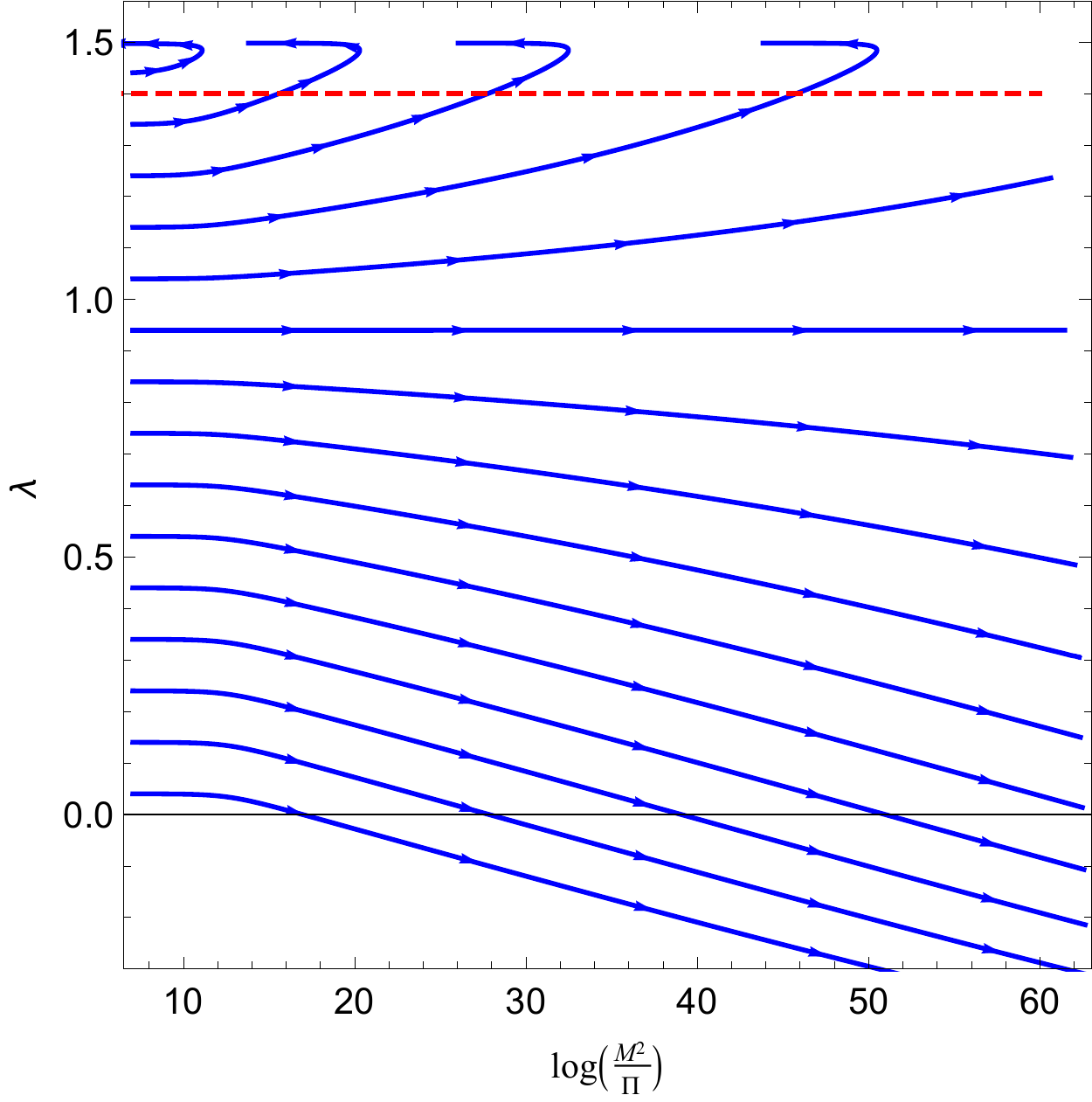}
  \caption{RG running with decreasing $t$ in the $\lambda$ - $\log(M^2/|\Pi|)$ plane and in the symmetric phase. 
The position of UV fixed point is now pushed to $\infty$ along the $x$ axis. 
The dashed line indicates that the solution at a fixed value of $\lambda_\Lambda$ while approaching $\Pi=0$ forces the system to run on trajectories closer and closer to the trajectory originating from $(\lambda_{UV}^*, M_{rUV}^{*2}$).}
  \label{flow-vs-M2Pi}
\end{figure}

This qualitative analysis can be followed more directly in 
Fig.\ref{flow-vs-M2Pi}, where we map the renormalization
trajectories in the $\lambda-\log(M^2/|\Pi|)$ plane. In this figure
the UV fixed point is shifted to infinity. One easily recognizes that the process which
increases $M^2_\Lambda/|\Pi_\Lambda|$ with $\lambda_{\Lambda}$ kept fixed, singles out
RG trajectories closer and closer to the renormalized trajectory originating from the point $(\lambda^*_{UV},\log(M_{UV}^{*2}/|\Pi_{UV}^*|)=\infty)$.

For a quantitative analysis, we will assume that the above process is
governed by a single UV fixed point. We introduce the usual fixed
point scaling laws,
\begin{equation}
  \delta\lambda = \lambda-\lambda_{UV}^* = (C_1 k)^{2\zeta},\qquad
  \delta M^2 = M_r^2-M_{rUV}^{*2} = (C_2 k)^{-\nu'}.
\end{equation}
 The powers $\zeta,\nu^\prime$ can be found by linearizing the LPA-form of the RGE's (\ref{DIMLESS-RGE-SYM}) around the fixed point (\ref{UV-FP-eq}). Expanding the eigenvalues to the same order in the small quantity $\lambda_{UV}^*v_d/(1+M_{rUV}^{*2})^3$, the expressions for the scaling powers are the following:
\be
2\zeta\approx\frac{1}{2\pi^2}\sqrt{\frac{5}{2}}h^2,\qquad \nu^\prime\approx 2-\frac{2}{5}\zeta.
\ee

We introduce
\begin{equation}
  \rho=\frac{M_r^2}{\Pi_r}.
\end{equation}
Near the fixed point, at large values of $\rho$ we find
\begin{equation}
  \rho \approx \frac{M_r^{*2}}{C/k^2} \sim k^2.
\end{equation}
Therefore, near the UV fixed point we can write
\begin{equation}
  \delta\lambda(k) = C \rho^\zeta(k).
\end{equation}
For smaller values of $\rho$, we extend its functional form as
\begin{equation}
  \delta\lambda(k) = C (\rho(k)+\rho_0)^{\zeta(k)}.
\end{equation}
In the IR limit, $\rho$ runs into a
constant (actually a negative value, due to SSB), and $\delta \lambda$
also reaches a constant value. Therefore, we have
\begin{equation}
  \delta\lambda_{IR} = \delta\lambda_{UV}
  \frac{(\rho_{IR}+\rho_0)^{\zeta_0}}{(\rho_{UV}+\rho_0)^{\zeta}}.
\end{equation}
This means that for large values of $\rho_{UV}$
\begin{equation}
  \lambda_{IR} = \lambda_{IR}^* + \kappa \rho_{UV}^{-\zeta}.
\end{equation}
This suggests a power-law approach to the limiting infrared value as was found in the numerical examples of the previous section. The above approximate expression for $\zeta$ leads to $\zeta(h=173/246)=0.0198$, which agrees satisfactorily with the numerically found value of the power for this case. For the second case ($h=3$), the agreement is worse, probably because it lies beyond the perturbative regime. The increasing tendency with $h^2$ is qualitatively confirmed.

\section{Effect of wave function renormalization ($LPA^\prime$)}
\label{wfr-effect}
The qualitatively new effect arising from taking into account anomalous dimensions is the nontrivial evolution of the Yukawa coupling, which suppresses the existence of the interacting LPA-fixedpoint in the ultraviolet. As one can see from combining (\ref{DIMLESS-RGE-SYM}) with (\ref{anom-dim-SYM}) in the neighborhood of the fixed point determined in $LPA$, the variation of the Yukawa coupling is well described as 
\be
h_r^2(t)=\frac{h_r^2(t_0)}{1-2h_r^2(t_0)C(t-t_0)},\qquad C\approx v_4\left(4+\frac{2}{1+M_{rUV}^{*2}}\right).
\ee
\begin{figure}[htbp]
  \centering
  \includegraphics[width=6cm]{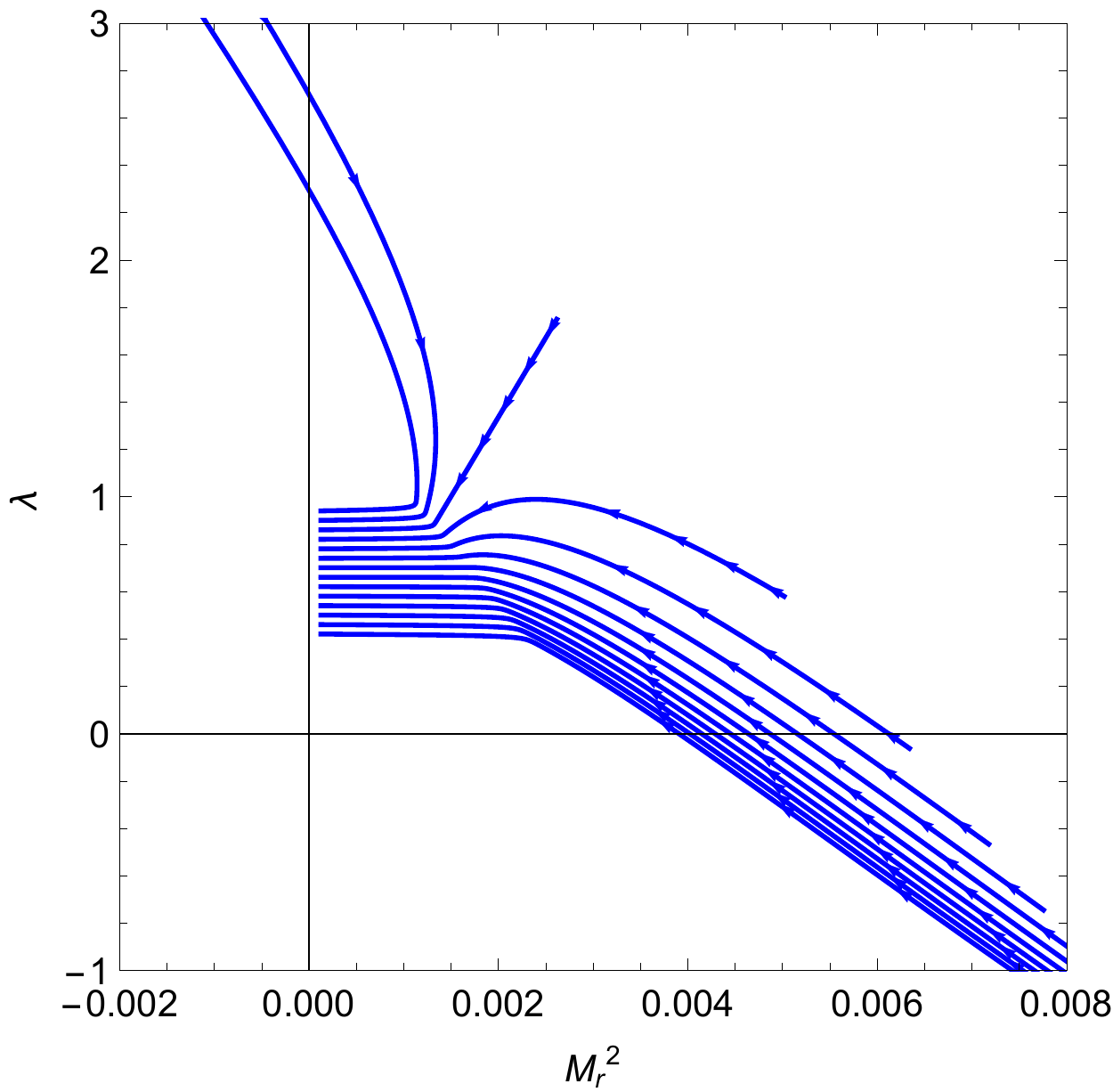}
\includegraphics[width=6cm]{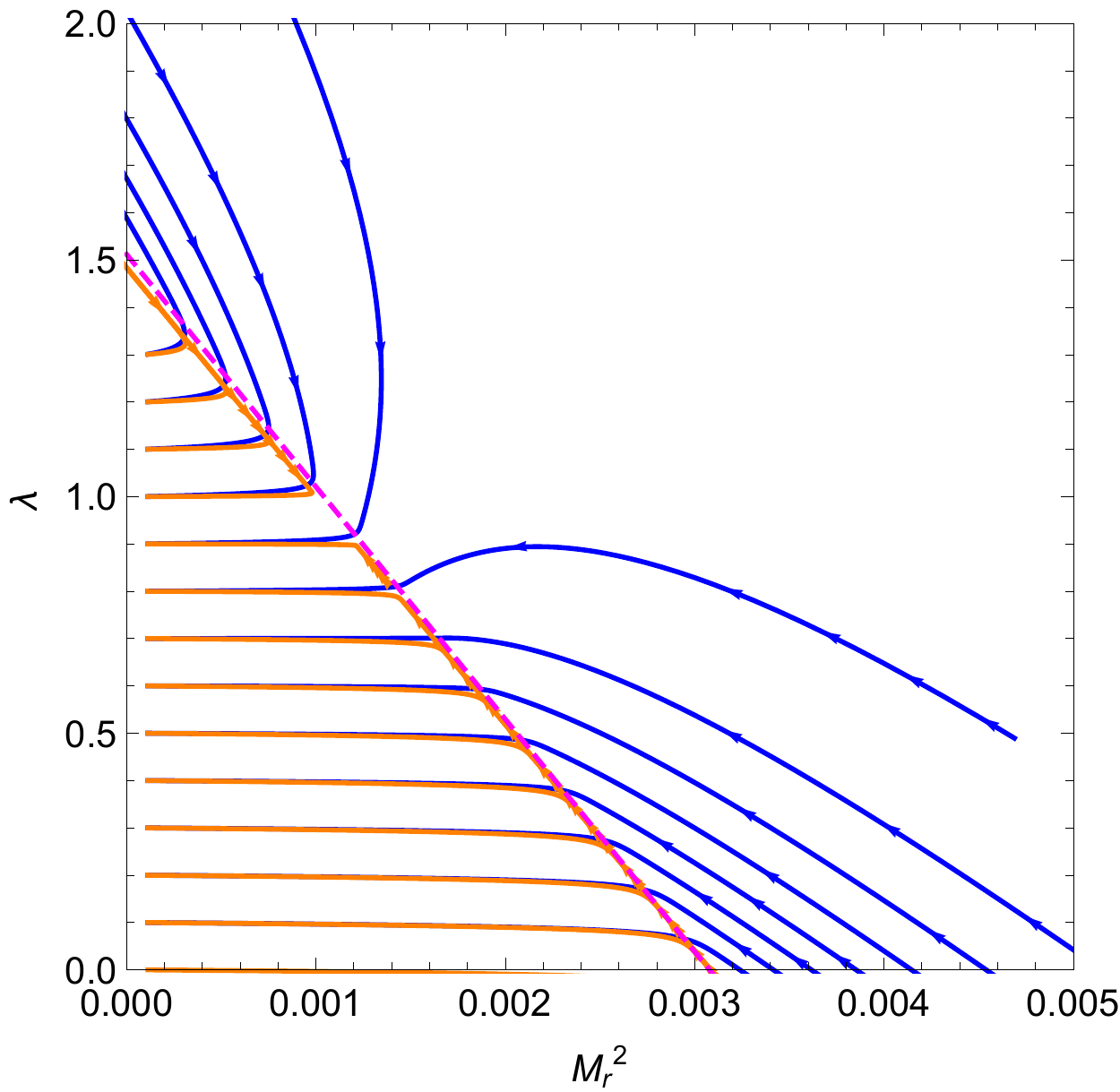}
  \caption{Flow diagram in the $\lambda-M_r^2$ space. All trajectories start with $h_r(\Lambda)=0.7$. There is no fixed point in $LPA^\prime$, still the neighborhood of the partial ($LPA$) fixed-point attracts the trajectories from the whole positive quarter.
{\it Left:} The trajectory separating the regions of UV instability from that of trajectories ending in a Landau singularity is clearly recognizable. {\it Right:} Magnified figure allows the comparison of the RG trajectories derived in $LPA$ (orange) and $LPA^\prime$ (blue). It emphasizes the close coincidence of the LPA-phase-separatrix (dashed red straight line) with the set of points where a sharp change in the character of the $LPA^\prime$ flow occurs. The arrows lead the eyes along the evolution towards infrared.}
  \label{fig:flowWFR}
\end{figure}

This logarithmic variation is rather slow compared to the powerlike scaling of the other couplings (in particular, since $|C|< 0.02$). Therefore one expects an "adiabatic" adaptation of the RG flow found in $LPA$ to the actual value of $h_r(t)$, which means that the flow described for fixed $h_r(\Lambda)$ in the previous section suffers only smooth deformation as $h_r(t)$ changes.

This argument is valid near the $LPA$-fixed-point. More generally one compares Fig.\ref{fig:flow1} to the flow diagram determined with $LPA^\prime$, e.g. Fig.\ref{fig:flowWFR}. The figure displays the flow for $|\Pi_\Lambda|/M_\Lambda^2=10^{-4}$, but its characteristic features do not change for weakening relative strength of the explicit symmetry breaking. Similar figures arise when $h_r(\Lambda)$ is varied.

Since there is no fixed point, the flow lines leading to the broken symmetry phase cover the complete positive quarter of the $\lambda_r-M_r^2$ plane. The most noticeable structural feature of the flow is a well recognizable remnant of the LPA separatrix as an almost straight line along which the RG trajectories go through a sharp (though continuous) change (see the right-hand plot in Fig.\ref{fig:flowWFR}). Left from this line the RG trajectories essentially coincide with those found in $LPA$.

\begin{figure}[htbp]
  \centering
  \includegraphics[width=6cm]{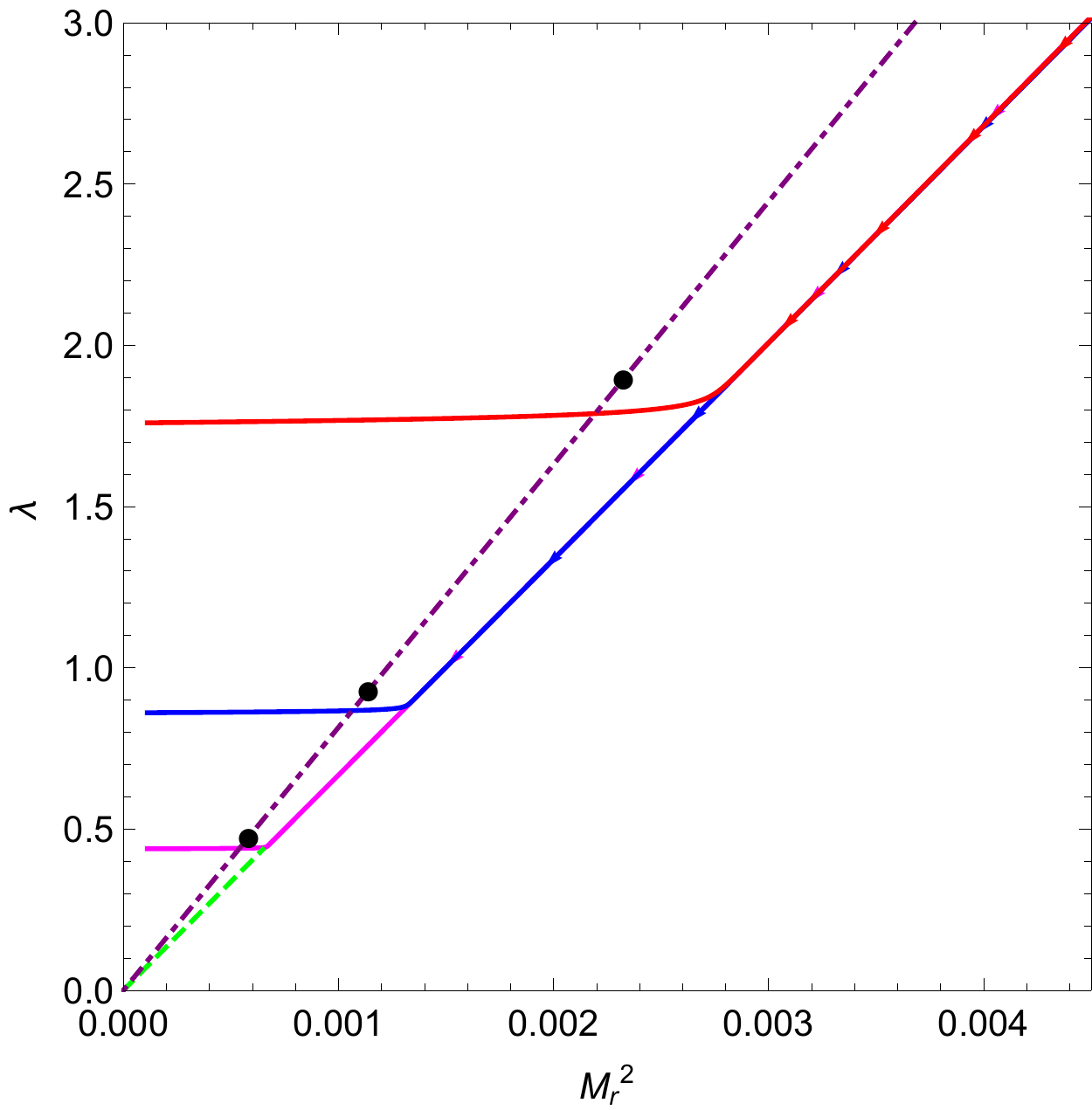}
  \caption{The trajectories taking the same $M^2=2|\Pi|$ and $h_r^2$ values, but different $\lambda_r$ at the critical surface  reach the "neutral" line in the $\lambda - M_r^2$ plane at different $h_r$. Next to the branching points also the LPA fixed points corresponding to these $h_r$ values are displayed.  Points of the partial LPA fixed line (see the dash-dotted line) as well as of the "neutral" line are labeled by the value of the Yukawa coupling. The two lines nearly coincide for lower values of $h_r$. }
  \label{fig:WFRflow_separator}
\end{figure}

With the help of a certain $\lambda_s^\prime$ one divides on the $\lambda_r-M_r^2$ plane the RG trajectories into two distinct classes. The trajectories which at criticality have $\lambda_r(M_r^2=M_{r,crit}^2)<\lambda_s^\prime(h_r(\Lambda))$ cross in the symmetric phase at some momentum scale through $\lambda_r=0$, defining the scale of instability. Trajectories which have on the critical surface 
$\lambda_r(M_r^2=M_{r,crit}^2)>\lambda_s^\prime(h_r(\Lambda))$ eventually reach a Landau singularity. There is a borderline trajectory, clearly recognizable on the left of Fig.\ref{fig:flowWFR}, separating these classes. This "neutral" trajectory is selected by choosing $\lambda_r(M_r^2=M_{r,crit}^2)=\lambda_s^\prime(h_r(\Lambda))$. 
This trajectory displays a linear $\lambda_r(M_r^2)$ dependence towards ultraviolet. The sequence of fixed points found in $LPA$ also form a straight line, only with slightly different slope as one sees in Fig.\ref{fig:WFRflow_separator}. The moderate shift is the result of the "adiabatic" deformation of the trajectories.

The predetermined $m_\psi^2/m_G^2$ and $\sqrt{2}m_\psi/u=h_r(k=0)$ values select some $h_r(\Lambda)$ and $|\Pi_r(\Lambda)|$.The "adiabatic" nature of the RG-flow deformation allows us to find this initial values rather close to the $LPA$ values. When one decreases $|\Pi_\Lambda|/M_\Lambda^2$ one finds the appropriate initial couplings just lying somewhat farther into UV along the $LPA^\prime$ RG trajectory.

The focusing array of RG trajectories is attracted by the neighborhood of the partial ($LPA$) fixed line labeled in the $\lambda_r,\Pi_r,M_r^2$ subspace by $h_\Lambda^2$. For $\Pi_\Lambda/M_\Lambda^2=10^{-2}-10^{-4}$ the initial values are rather close to the former separatrix. The effect of the wave function renormalization is minimal ($\eta_i$ are both rather small) along the trajectory. Therefore the $LPA^\prime$ predictions for the mass ratio $m_G/m_{hb}$ are almost indistinguishable from the $LPA$ results. This can be seen clearly in Fig.\ref{explicit-SB-dependence} for $\lambda_\Lambda=0.5, 1.4$. The mass ratio belonging to $\lambda_s^\prime$ is slightly different relative to the $LPA$ case, as one can see comparing the two horizontal lines in the same figure.

Eventually, it can be argued that similar phenomenon can be observed for the ratio $m_G/m_{hb}$ when weakening the relative strength of the explicit symmetry breaking what has been described above in LPA. Let us keep $\lambda_r(k=\Lambda)$ fixed at some value above $\lambda_s^\prime(M_r^2=M_{r,crit}^{2})$. With decreasing $|\Pi_\Lambda|/M_\Lambda^2$  one moves first over to trajectories closer to the trajectory of $\lambda_s^\prime$ (the "neutral" trajectory). This means that first one approaches the mass ratio belonging to $\lambda_s^\prime$. In contrast to LPA, now one will cross the "neutral" trajectory and the corresponding $\lambda_r(M_{r, crit})$ moves away from this line, further increasing the predicted bosonic mass ratio. However, each trajectory in this region reaches a maximum at some $M_{r,max}^2$. The chosen $\lambda_r(\Lambda)$ touches upon this turning point belonging to a certain RG trajectory. When passing beyond $M_{r,max}^2$ one crosses again trajectories lying closer and closer to the "neutral" one. When one fixes an initial $\lambda_\Lambda$ below $\lambda_s^\prime$, one repeats the second half of this analysis. When the starting  trajectory has no maximum at all, then with decreasing 
$|\Pi_\Lambda|/M_{r,\Lambda}$ one approaches monotonically to the ratio belonging to the "neutral" trajectory. In view of this qualitative analysis one expects that the ratio of the scalar masses at $k=0$ will approach in the limit of vanishing explicit symmetry breaking the set of IR couplings reached along the continuation of the "neutral" trajectory into the broken symmetry phase. The only change relative to $LPA$ is that the convergence is now nonmonotonic.

\section{Summary of the results and extensions}
\label{conclusion}

In this paper we have shown that in the local potential approximation of non-perturbative renormalisation group equations a $U_L(1)\times U_R(1)$ symmetric Yukawa theory with a quadratic explicit symmetry breaking term possesses a UV fixed line in the symmetric phase, characterized by vanishing strength of the explicit symmetry breaking. It is the value of the Yukawa coupling, an RG invariant in this approximation, which parametrizes the line of fixed points in the three-dimensional $(\lambda_\Lambda, h_\Lambda, M_\Lambda^2)$ coupling space. 

Although the Yukawa coupling starts to evolve logarithmically in the next order of the gradient expansion of the effective action, effacing in this way the partial fixed line, it was found that a slightly displaced neighborhood of the $LPA$ fixed line ($\lambda_r^*(h_\Lambda),M_r^{2*}(h_\Lambda)$) still attracts the RG trajectories from the far ultraviolet ($\lambda_r>\lambda_r^*(h_\Lambda), M_r^2>M_r^{2*}(h_\Lambda)$). In the region $\lambda_r<\lambda_r^*(h_\Lambda), M_r^2<M_r^{2*}(h_\Lambda)$ the RG flow of the $LPA$ and $LPA^\prime$ essentially coincide.

The mechanism behind this focusing effect is made explicit by comparing the RG flow in $LPA$ and $LPA^\prime$. It appears useful to summarize in this conclusion first the general features which lead to the existence of a partial UV fixed point and comment on the changes occurring to its effects in the next approximation steps at the end.

The UV fixed point was found in $LPA$ with a quartic local potential. The part of the RG flow was mapped out which crosses the critical surface, bordering the phase of broken $U_A(1)$ symmetry. The renormalized trajectory starting from the fixed point characterized by some fixed value of $h$ has in the symmetric phase a constant $\lambda=\lambda_s$. Above this value, flow lines of the trajectories arrive from the neighborhood of Landau poles, restricting the range of allowed momenta. Below the renormalized trajectory the physically allowed portion of the trajectories is restricted by the stability requirement $\lambda>0$.

An important feature is that trajectories passing near the unique fixed point of the quadratic symmetry breaking ($\Pi_r^*=0$), in the broken symmetry are able to sustain stable fermion-to-boson and boson-to-boson mass ratios not enforcing in the limiting case the validity of Goldstone's theorem. 

The key points needed for the establishment of such an $LPA$ fixed line in a certain theory are the following:
\begin{itemize}
\item{The RG-evolution of the explicit quadratic symmetry breaking parameter $\Pi$ is driven by itself in the symmetric phase, resulting from a convenient cancellation among the contributions from the bosons as well, as a complete cancellation of the fermionic contribution.}
\item{The rate of the RG-evolution of the Yukawa-coupling  $h$ is also proportional to the explicit symmetry breaking parameter $\Pi$, resulting from a convenient cancellation among the bosonic contributions.}
\item{The existence of the interacting $LPA$ fixed point ($\lambda_{UV}^*, M_{rUV}^*$) is made possible by the compensation of the bosonic contributions (proportional to some power of the self-coupling $\lambda$) through the fermionic contribution (proportional to the non-running Yukawa coupling). It is a non-trivial requirement to find this fixed point in the physically allowed range.}
\end{itemize}

In order to demonstrate the generic nature of the construction leading to such a UV fixed line, we shortly demonstrate its existence in a well-known model with a larger symmetry group.
Let consider the $U_L(2)\times U_R(2)$ symmetric chiral "quark-meson" theory.  The bosonic fields of the model are written in a $2\times 2$ matrix with complex elements and transform as follows:
\be
U=\sigma+i\eta+(s_a+i\pi_a)\tau^a, \qquad U\rightarrow U_LUU_R^\dagger, \quad U_L\in SU(2)_L,\quad U_R\in SU(2)_R
\ee
The part of the action invariant under $U_L(2)\times U_R(2)$ is the following:
\be
\Gamma_{INV}=\int_x\left[\frac{1}{4}{\textrm{Tr}}\partial_\alpha U^\dagger \partial_\alpha U+\bar\psi\sls\partial\psi+V(\rho)+h(\bar\psi_L U\psi_R+\bar\psi_R U^\dagger \psi_L)\right],
\ee
with $\rho=(1/4){\textrm{Tr}}U^\dagger U$. The fermion field $\psi_{L/R}$ represents an $SU_{L/R}(2)$ doublet.
The quadratic explicit symmetry breaking is chosen in the following specific form, closed upon the RG evolution:
\be
\Gamma_{DSB}=\int_x\bar\Pi({\textrm{Tr}}U^2+{\textrm{Tr}}(U^\dagger)^2).
\ee

Assuming a nonzero background for the $\sigma$ field one quickly finds the expression for $\Gamma^{(2)}$ and one proceeds to the straightforward derivation of the RGE's, noting the fact that in a pure $\sigma$ background with the above explicit symmetry breaking $G_\pi=G_\eta$ in both phases and $G_s=G_\sigma$ in the symmetric phase. 
The RGE's for the $\sigma$ and $\pi$ fields in the symmetric phase ($V^\prime=M^2, V^{\prime\prime}=\lambda/3$) are the following:
\bea
\partial_t (M^2+\bar\Pi)&=&\frac{\lambda}{3}\hat\partial_t\int_q(3G_\sigma+2G_\pi)-8h^2\hat\partial_t\int_q\frac{1}{q^2_R},\nonumber\\
\partial_t(M^2-\bar\Pi)&=&\frac{\lambda}{3}\hat\partial_t\int_q(2G_\sigma+3G_\pi)-8h^2\hat\partial_t\int_q\frac{1}{q^2_R},
\eea
or
\bea
\partial_t M^2&=&\frac{5\lambda}{6}\hat\partial_t\int_q(G_\sigma+G_\pi)-8h^2\hat\partial_t\int_q\frac{1}{q_R^2},\nonumber\\
 \partial_t\bar\Pi&=&\frac{\lambda}{6}\hat\partial_t\int_q(G_\sigma-G_\pi)=-\frac{\lambda\bar\Pi}{3}\hat\partial_t\int_q\frac{1}{(q_R^2+M^2)^2-\bar\Pi^2}.
\eea
Clearly, the explicit symmetry breaking parameter is driven by itself and it has the fixed point $\bar\Pi^*=0$.

In the symmetric phase only fermions contribute to the RGE of the Yukawa coupling:
\be
\partial_th=-\frac{h^3}{3}\hat\partial_t\int_q\frac{1}{q_R^2}[G_\sigma-G_\eta+3(G_S-G_\pi)]=-\frac{4}{3}h^3\hat\partial_t\int_q\frac{1}{q_R^2}(G_\sigma-G_\pi),
\ee
which is proportional to $\bar\Pi$ by the difference $G_\sigma-G_\pi$ as was the case of the RGE (\ref{RGE-SB-2}) (compare with the first two points of the general scheme). Therefore, any chosen value of $h$ might represent a fixed point solution of $M_r^2$ and $\lambda$.
The RGE of $\lambda$ is structurally the same as was the case of the $U(1)$ symmetric theory,
\be
\partial_t\lambda=48h^4\hat\partial_t\int_q\frac{1}{q_R^4}-\frac{\lambda^2}{2}\hat\partial_t\int_q\left(G_s^2+\frac{4}{3}G_\pi^2+3G_\sigma^2\right).
\ee

Analyzing the RGE's one quickly finds the following approximate fixed point equations for any chosen value of $h^2$ valid in the ultraviolet under the condition of the negligibility of $M_r^2$ on the right-hand side of the RGE's:
\be
\Pi_{UV}^*=0, \qquad \frac{8}{3}\lambda_{UV}^{*2}=48h^4,\qquad 2M_{rUV}^{*2}=2v_d\left(8h^2-\frac{5}{3}\lambda_{UV}^*\right),
\ee
having the solution
\be
\Pi_{UV}^*=0,\qquad \lambda_{UV}^*=\sqrt{18}h^2,\qquad M_{rUV}^{*2}=\frac{1}{32\pi^2}(8-5\sqrt{2})h^2.
\ee
This solution is physical, since the bracket in the expression of $M_{rUV}^{*2}$ is positive. It is also compatible with the condition $M_{rUV}^{*2}<<1$, and the RG flow should be rather similar in its neigbhborhood to that described in detail for the $U(1)$ case.
The particle spectra in the broken symmetry phase is far from the phenomenologically interesting case of strong interactions; therefore,the interest of the quadratic explicit symmetry breaking is doubtful for the phenomenology of any realistic two-flavor quark-meson theory.

The generality of the constructional requirements makes it clear that including higher (perturbatively irrelevant) powers of the invariant $\rho$ into the potential $V(\rho)$ will only slightly influence the result of the analysis for moderate values of the Yukawa coupling. The upper bound on $h$ will depend on the algebraic structure of the model.

Nonzero anomalous dimensions (the so-called $LPA^\prime$) efface the fixed line. Since the logarithmic evolution of the Yukawa coupling in the region of the $LPA$ fixed line is parametrically slower than the scaling of the other couplings, some kind of "adiabatic" deformation of the RG flow found in $LPA$ has a chance to occur. The coefficient governing the logarithmic variation depends on the symmetry group of the theory; therefore, this expectation should be checked numerically case by case.    

Possible significance of theories, whose fully massive infrared spectra is dominated by the remnants of a partial UV fixed line in the limit of vanishing explicit symmetry breaking, for consistent ultraviolet completion should be investigated in the future systematically. It would be also of interest to reconstruct the features of this study from a purely fermionic version of this model (inverse Hubbard-Stratonovich transformation). Lattice realization of the RG evolution would provide a nonperturbative check of the findings of the present paper.

\section*{APPENDIX: EXPLICIT RENORMALIZATION GROUP EQUATIONS}

The RGE's were derived with the linear (optimized) cutoff functions of Litim \cite{litim01}
\be
R_B(q)=q^2\left(\frac{k^2}{q^2}-1\right)\Theta\left(\frac{k^2}{q^2}-1\right),\qquad
R_F(q)=\sls{q}\left(\frac{k}{q}-1\right)\Theta\left(\frac{k^2}{q^2}-1\right).
\ee
In the symmetric phase the following basic relations are used:
\bea
&&
\hat\partial_t\int_q\frac{1}{(Z_\psi^2q^2_R+m_\psi^2)^n}=-2nv_dk^{d+2}\frac{1}{(Z_\psi^2k^2+m_\psi^2)^{n+1}}\left(1-\frac{\eta_\psi}{d+1}\right),\nonumber\\
&&
\hat\partial_t\int_q\frac{1}{((Z_\phi q^2_R+M^2)^2-4\Pi^2)^n}=-4nv_dk^{d+2}\frac{Z_\phi k^2+M^2}{((Z_\phi k^2+M^2)^2-4\Pi^2)^{n+1}}\left(1-\frac{\eta_\phi}{d+2}\right),\nonumber\\
\eea
where $v_d=S_d/(d(2\pi)^d)$ with $S_d$ denoting the surface of the $d$-dimensional unit sphere. In the broken symmetry phase the following general relation is used in the derivation of the RGE's:
\bea
&&
\hat\partial_t\int_qG_G^l(q_R^2)G_{hb}^n(q_R^2)=-2v_dk^{d+2}G_G^l(k^2)G_{hb}^n(k^2)(lG_G(k^2)+nG_{hb}(k^2))
\left(1-\frac{\eta_\phi}{d+2}\right),
\nonumber\\
&&
G_G(q_R^2)=\frac{1}{Z_\Phi q_R^2+m_G^2},\qquad G_{hb}(q_R^2)=\frac{1}{Z_\Phi q_R^2+m_{hb}^2}.
\eea

In the symmetric phase and in four dimensions [$v_4=(32\pi^2)^{-1}$], the following set is obtained for the couplings defining the ansatz for the effective action as the sum of $\Gamma_{INV}$ and $\Gamma_{DSB}$, when one takes into account wave function renormalization and a quartic scalar potential:
\bea
&&
\partial_t\Pi_r+(2-\eta_\phi)\Pi_r=\frac{4\lambda_r\Pi_rv_4}{3}\frac{1+M_r^2}{((1+M_r^2)^2-4\Pi_r^2)^2}\left(1-\frac{\eta_\phi}{6}\right),\nonumber\\
&&
\partial_tM_r^2+(2-\eta_\phi)M_r^2=4h_r^2v_4\left(1-\frac{\eta_\psi}{5}\right)-
\frac{4\lambda_rv_4}{3}\frac{(1+M_r^2)^2+4\Pi_r^2}{((1+M_r^2)^2-4\Pi_r^2)^2}\left(1-\frac{\eta_\phi}{6}\right),\nonumber\\
&&
\partial_t\lambda_r-2\eta_\phi\lambda_r=-24h_r^4v_4\left(1-\frac{\eta_\psi}{5}\right)+
\frac{4\lambda_r^2v_d(1+M_r^2)}{3((1+M_r^2)^2-4\Pi_r^2)^3}\left(5(1+M_r^2)^2+52\Pi_r^2\right)\left(1-\frac{\eta_\phi}{6}\right),\nonumber\\
&&
\partial_t h_r^2-(\eta_\phi+2\eta_\psi)h_r^2=-\frac{4\Pi_rh_r^4v_4}{(1+M_r^2)^2-4\Pi_r^2}\left[\frac{2(1+M_r^2)}{(1+M_r^2)^2-4\Pi_r^2}\left(1-\frac{\eta_\phi}{6}\right)+1-\frac{\eta_\psi}{5}\right].
\label{DIMLESS-RGE-SYM}
\eea
The algebraic equations determining the anomalous dimensions look like
\be
\eta_{\psi}=h_r^2v_4\frac{(1+M_r^2)^2+4\Pi_r^2}{((1+M_r^2)^2-4\Pi_r^2)^2}\left(1-\frac{\eta_\phi}{5}\right),\qquad \eta_\phi=h_r^2v_4(4-\eta_\psi).
\label{anom-dim-SYM}
\ee 

A test of the robustness of the conclusions obtained in LPA with a quartic potential is performed by investigating the effect of the next term of the expansion of the scalar potential $\lambda_3 (\Phi^*\Phi)^3\equiv\lambda_3\rho^3$. The determinant of $\Gamma_B^{(2)}$ supplemented with this term in $LPA$ has the following expression:
\be
\Delta(q_R^2,\Phi,\Phi^*)=\left(q_R^2+M^2+\frac{2\lambda}{3}\rho+9\lambda_3\rho^2\right)^2-\left(2\Pi+\frac{\lambda}{3}\Phi^{*2}+6\lambda_3\rho\Phi^{*2}\right)\left(2\Pi+\frac{\lambda}{3}\Phi^{2}+6\lambda_3\rho\Phi^{2}\right).
\ee
It affects the bosonic contribution to $\Gamma_{\Phi^2\Phi^{*2}}$ and $\Gamma_{\Phi^3\Phi^{*3}}$. The derivatives of $\Delta$ with respect to $\Phi$ and $\Phi^*$ are denoted as
\be
\Delta_{nm}=\frac{\delta^{n+m}\Delta}{\delta\Phi^n\delta\Phi^{*m}}.
\ee
The nonvanishing derivatives at the origin are the following:
\bea
&\displaystyle
\bar \Delta_{11}=\frac{4\lambda}{3}(q^2+M^2),\qquad 
\bar \Delta_{22}=4\left(\frac{\lambda^2}{3}+18\lambda_3(q^2+M^2)\right),\qquad \bar \Delta_{33}=288\lambda\lambda_3,\nonumber\\
&\displaystyle
\bar \Delta_{20}=\bar \Delta_{02}=-\frac{4\lambda}{3}\Pi,\qquad \bar \Delta_{31}=\bar \Delta_{13}=-72\lambda_3\Pi.
\label{non-zero-der}
\eea
One checks readily that the tadpole integral computed with the 6-leg term contributes to the 
right-hand side of $\Gamma_{\Phi^2\Phi^{*2}}$, that is to the running of $2\lambda/3$,
\be
18\lambda_3\hat\partial_t\int \left(G_G(q_R^2)+G_{hb}(q_R^2)\right),\qquad G_G=\frac{1}{q_R^2+M^2-2\Pi},\qquad G_{hb}=\frac{1}{q_R^2+M^2+2\Pi}.
\ee
Taking further two derivatives, one finds the bosonic contribution to the RGE of  $\lambda_3$ (after dividing both sides by 36),
\be
\hat\partial_t\frac{\delta^6}{\delta\Phi^3\delta\Phi^{*3}}\frac{1}{72}\int_q\log\Delta(q_R^2,\Phi,\Phi^*)=
\frac{1}{72}\hat\partial_t\int_q\Biggl[\frac{\Delta_{33}}{\Delta}
-\frac{3}{\Delta^2}\left(3\Delta_{11}\Delta_{22}+2\Delta_{20}\Delta_{31}\right)
+\frac{6}{\Delta^3} \left(2\Delta^3_{11}+3\Delta^2_{20}\Delta_{11}\right)\Biggr].
\label{bosonic-contrib-lambda3}
\ee
The fermionic contribution to $\partial_t\lambda_3$ is given with help of the massless fermion propagator $G_F$ and the chiral projectors $P_\pm$,
\bea
-\frac{1}{36}\frac{\delta^6}{\delta\Phi^3\delta\Phi^{*3}}\hat\partial_t{\textrm{Trlog}}\Gamma_F^{(2)}&=&\frac{1}{3}h^6{\textrm{Tr}}(G_FP_+G_FP_-G_FP_+G_FP_-G_FP_+G_FP_-)\nonumber\\
&
=&-\frac{2}{3}h^6\int_q\hat\partial_t\frac{1}{(P_F^2(q)q^2)^3}.
\label{fermionic-contrib-lambda3}
\eea
In the broken symmetry phase, it is convenient to introduce a separate notation for the scaled propagators in $LPA^\prime$,
\be
d_G=\frac{1}{1+\mu_G^2},\qquad  d_{hb}=\frac{1}{1+\mu_{hb}^2},\qquad
d_\psi=\frac{1}{1+\mu_\psi^2}.
\ee
One writes for the Yukawa coupling and the scaled masses after a lengthy but straightforward computation the following RGE's:
\be
\mu^2_{G}=\frac{m^2_G}{Z_\phi k^2},\qquad \mu^2_{hb}=\frac{m^2_{hb}}{Z_\phi k^2},\qquad \mu^2_{\psi}=\frac{m_\psi^2}{Z_\psi^2 k^2}
\ee
\bea
&&\frac{3}{4}\left(\partial_t\mu_G^2+(2-\eta_\phi)\mu_G^2\right)=4h_r^2v_4d_\psi^2\left(1-\frac{\eta_\psi}{5}\right)\nonumber\\
&&~~~~~~
~~~~~~~~~~~~-\frac{\lambda_r}{3}v_4\left(1-\frac{\eta_\phi}{6}\right)\left[2d_{hb}^2+2d_G^2-(\mu_{hb}^2+\mu_G^2)(2+\mu_{hb}^2+\mu_G^2)d_{hb}^2d_G^2\right],\nonumber\\
&&\partial_t\mu_{hb}^2+(2-\eta_\phi)\mu_{hb}^2=4h_r^2v_4(1-\mu_\psi^2)d_\psi^3
\left(1-\frac{\eta_\psi}{5}\right)\nonumber\\
&&~~~-\frac{4\lambda_r}{3}v_4\left(1-\frac{\eta_\phi}{6}\right)
\Bigl[d_G^2+\frac{3}{4}(\mu_{hb}^2+\mu_G^2)d_Gd_{hb}(d_G+d_{hb})
-2\mu_{hb}^2(d_{hb}^3+d_G^3+d_{hb}d_G(d_{hb}+d_G))\nonumber\\
&&~~~~~~~~~~~~~~~~~~~~~+4\mu_{hb}^2(\mu_{hb}^2+\mu_G^2)(d_{hb}^3d_G+d_{hb}d_G^3+d_{hb}^2d_G^2)
-\frac{\mu_{hb}^2}{2}(\mu_{hb}^2+\mu_G^2)^2d_{hb}^2d_G^2(d_{hb}+d_G))\Bigr],
\nonumber\\
&&
\partial_th_r^2-(\eta_\phi+2\eta_\psi)h_r^2
\nonumber\\
&&=2h_r^4v_4\Biggl[\left(1-\frac{\eta_\psi}{5}\right)d_\psi^2\left[\mu_{hb}^2\left(\frac{1}{2}d_G^2-\frac{3}{2}d_{hb}^2-d_Gd_{hb}\right)
+\mu_G^2d_Gd_{hb}+4\mu_\psi^2(\mu_{hb}^2-\mu_G^2)d_Gd_{hb}d_\psi\right]\nonumber\\
&&~~~~~~~~~~+\left(1-\frac{\eta_\phi}{6}\right)d_\psi\left[(\mu_G^2-\mu_{hb}^2)(1-2\mu_\psi^2d_\psi)d_Gd_{hb}(d_G+d_{hb})+\mu_{hb}^2(d_G^3-3d_{hb}^3)\right]\Biggr].
\eea
The equations of the anomalous dimensions smoothly join those valid in the symmetric phase:
\bea
&&\eta_\phi=\frac{2\lambda_r}{3}v_4\mu_{hb}^2\left(2d_G^2d_{hb}^2+\frac{1}{4}(d_G^2+d_{hb}^2)^2\right)+h_r^2v_4(1-\mu_\psi^2)d_\psi^3\left(2-\eta_\psi+2d_\psi\right),\nonumber\\
&& \eta_\psi=\frac{1}{2}h_r^2v_4\left(1-\frac{\eta_\phi}{5}\right)d_\psi(d_G^2+d_{hb}^2).
\eea

The fermionic contribution to the evolution of $2\lambda/3$ is given by the fermionic quadrangle diagram:
\be
\frac{2}{3}\partial_t^{F}\lambda=4h^4\hat\partial_t\int_pG_\psi^2[1-4m_\psi^2G_\psi+m_\psi^4G_\psi^2],\qquad G_\psi=\frac{1}{Z_\psi^2P_F^2(q^2)q^2+m_\psi^2}.
\ee
The bosonic contribution is much more cumbersome therefore only a semiexplicit expression is given here with help of appropriate $\Phi$ and $\Phi^*$ derivatives of the bosonic determinant 
\be
\Delta(q_R^2, \Phi,\Phi^*)=\left[Z_\Phi q_R^2+\frac{\lambda}{3}\left(2\Phi\Phi^*-\frac{v^2}{2}\right)\right]^2-\left(2\Pi+\frac{\lambda}{3}\Phi^2\right)\left(2\Pi+\frac{\lambda}{3}\Phi^{*2}\right).
\ee
The bosonic contribution to $2\partial_t\lambda /3$ takes the following form:
\bea
\frac{2}{3}\partial_t^B\lambda&=&\frac{1}{2}\hat\partial_t\int_p\Biggl\{\frac{\Delta_{22}}{\Delta}-
\frac{1}{\Delta^2}\left[2\Delta_{12}\Delta_{10}+
2\Delta_{21}\Delta_{01}+2\Delta_{11}^2+\Delta_{20}\Delta_{02}\right]\nonumber\\
&&
+\frac{2}{\Delta^3}\left[\Delta_{20}\Delta_{01}^2+\Delta_{02}\Delta_{10}^2+4\Delta_{11}\Delta_{10}\Delta_{01}\right]-\frac{6\Delta_{10}^2\Delta_{01}^2}{\Delta^4}\Biggr\}.
\eea
One finds the explicit expression when each quantity appearing on the right-hand side is evaluated at $\Phi=\Phi^*=u/\sqrt{2}$. One has to substitute the following nonzero quantities, when a quartic potential is being used:
\bea
&&
\bar\Delta_{10}=\bar\Delta_{01}=\frac{2\lambda}{3}\frac{u}{\sqrt{2}}\left[2Z_\Phi q_R^2+\frac{\lambda}{2}u^2-\frac{\lambda}{3}v^2-2\Pi\right],\nonumber\\
&&
\bar\Delta_{11}=\frac{2\lambda}{3}\left[2Z_\Phi q_R^2+\lambda u^2-\frac{\lambda}{3}v^2\right],\qquad \bar\Delta_{22}=\frac{4\lambda^2}{3},\nonumber\\
&&
\bar\Delta_{20}=\Delta_{02}=\frac{\lambda}{3}(\lambda u^2-4\Pi),\qquad
\bar\Delta_{21}=\Delta_{12}=\frac{4\lambda^2}{3}\frac{u}{\sqrt{2}},\nonumber\\
&&
\bar\Delta=\left(Z_\Phi q^2_R+\frac{\lambda}{3}u^2\right)(Z_\Phi q^2_R-4\Pi).
\eea
The relevant combinations can be expressed through the bosonic masses, taking into account that at $\Phi=\Phi^*=u/\sqrt{2}$, $\Delta_{nm}=\Delta_{mn}$,
\bea
&&
\Delta =(q_R^2+m_{hb}^2)(q_R^2+m_G^2),\qquad \Delta_{11}=\frac{2\lambda}{3}\left(G_{hb}^{-1}+G_G^{-1}+m_{hb}^2\right),\qquad \Delta_{22}=\frac{4\lambda^2}{3},\nonumber\\
&&
\Delta_{01}^2=\frac{\lambda}{6}m_{hb}^2\left(G_{hb}^{-1}+3G_G^{-1}\right)^2,\qquad \Delta_{01}\Delta_{12}=\frac{2\lambda^2}{3}m_{hb}^2\left(G_{hb}^{-1}+3G_G^{-1}\right),\qquad \Delta_{02}=\frac{\lambda}{3}(3m_{hb}^2+m_G^2).
\eea
Using these in the bosonic contribution to the RGE of $\lambda$, one can express the integrand in powers $G_{hb}^lG_G^k$ to which one easily applies the $\hat\partial_t$ operation.

\section*{Acknowledgement}
This research was supported by the Hungarian Research Fund under Contracts No. K104292 and No. K123815. Valuable suggestions of an unknown Referee prompting considerable expansion of the scope of the first version of this communication, are thankfully acknowledged.

\end{document}